\documentclass[10pt,letterpaper]{article}
\usepackage[headheight=28pt,top=0.85in,left=2.75in,footskip=0.75in]{geometry}

\usepackage{changepage}

\usepackage[utf8]{inputenc}

\usepackage{textcomp,marvosym}

\usepackage{fixltx2e}

\usepackage{amsmath,amssymb}

\usepackage{cite}

\usepackage{nameref,hyperref}

\usepackage{microtype}
\DisableLigatures[f]{encoding = *, family = * }

\usepackage{rotating}

\usepackage{setspace} 
\raggedright
\usepackage[aboveskip=1pt,labelfont=bf,labelsep=period,justification=raggedright,singlelinecheck=off]{caption}

\makeatletter
\renewcommand{\@biblabel}[1]{\quad#1.}
\makeatother

\date{}

\usepackage{lastpage,fancyhdr,graphicx}

\usepackage{epstopdf}
\fancyheadoffset[L]{2.25in}
\fancyfootoffset[L]{2.25in}
\newcommand{\ihmmunealign}{\textsf{iHMMunealign}}
\newcommand{\imgt}{\textsf{IMGT}}
\newcommand{\partis}{\textsf{partis}}
\newcommand{\ham}{\textsf{ham}}
\newcommand{\igblast}{\textsf{IgBLAST}}
\newcommand{\soda}{\textsf{SODA}}
\newcommand{\sodatwo}{\textsf{SODA2}}
\newcommand{\stochhmm}{\textsf{StochHMM}}
\newcommand{\hmmoc}{\textsf{HMMoC}}
\newcommand{\ighutil}{\textsf{ighutil}}

\newcommand{\vdj}{VDJ}

\newcommand{\forarxiv}[1]{#1}
\newcommand{\notforarxiv}[1]{}

\newcommand{\argmax}{\operatorname{argmax}}

\newcommand{\beginsupplement}{%
        \setcounter{table}{0}
        \renewcommand{\thetable}{S\arabic{table}}%
        \setcounter{figure}{0}
        \renewcommand{\thefigure}{S\arabic{figure}}%
     }

\newcommand{\TABLEhamPerformance}{\
{\small
\begin{table}[ht]
\centering
\caption{\bf HMM compiler performance comparison}
\begin{tabular}{|l|c|c|c|}
  \hline
  method & seq length (x$10^6$) & user time (sec)   & max memory (MB)  \\ \hline
  \hmmoc &  1                   & $0.57 \pm 0.01$   & $18.01 \pm 0.01$ \\ \hline
  \ham   &  1                   & $0.54 \pm 0.01$   & $12.65 \pm 0.0 $ \\ \hline
  \hmmoc &  10                  & $5.05 \pm 0.07$   & $168.7 \pm 0.1 $ \\ \hline
  \ham   &  10                  & $4.77 \pm 0.05$   & $110.6 \pm 0.1 $ \\ \hline
\end{tabular}
\caption*{\
  Efficiency of \ham\ compared to code generated by \hmmoc\ for the ``occasionally dishonest casino'' from~\cite{Durbin1998-uq}.
  Elapsed CPU time and memory usage (user time and maximum resident set size from the Unix \texttt{time} command) for the Forward probability calculation are shown for sequences of length 1 million (mean of 300 runs) and 10 million (mean of 30 runs).
  These estimates do not include the extra time for code generation and compilation which is necessary in \hmmoc.
  Uncertainties are the standard error on the mean.
}
\label{TABLEhamPerformance}
\end{table}}
}

\newcommand{\FIGaRecoEvent}{\
  \begin{figure}[ht]
    \forarxiv{\includegraphics[width=3.5in]{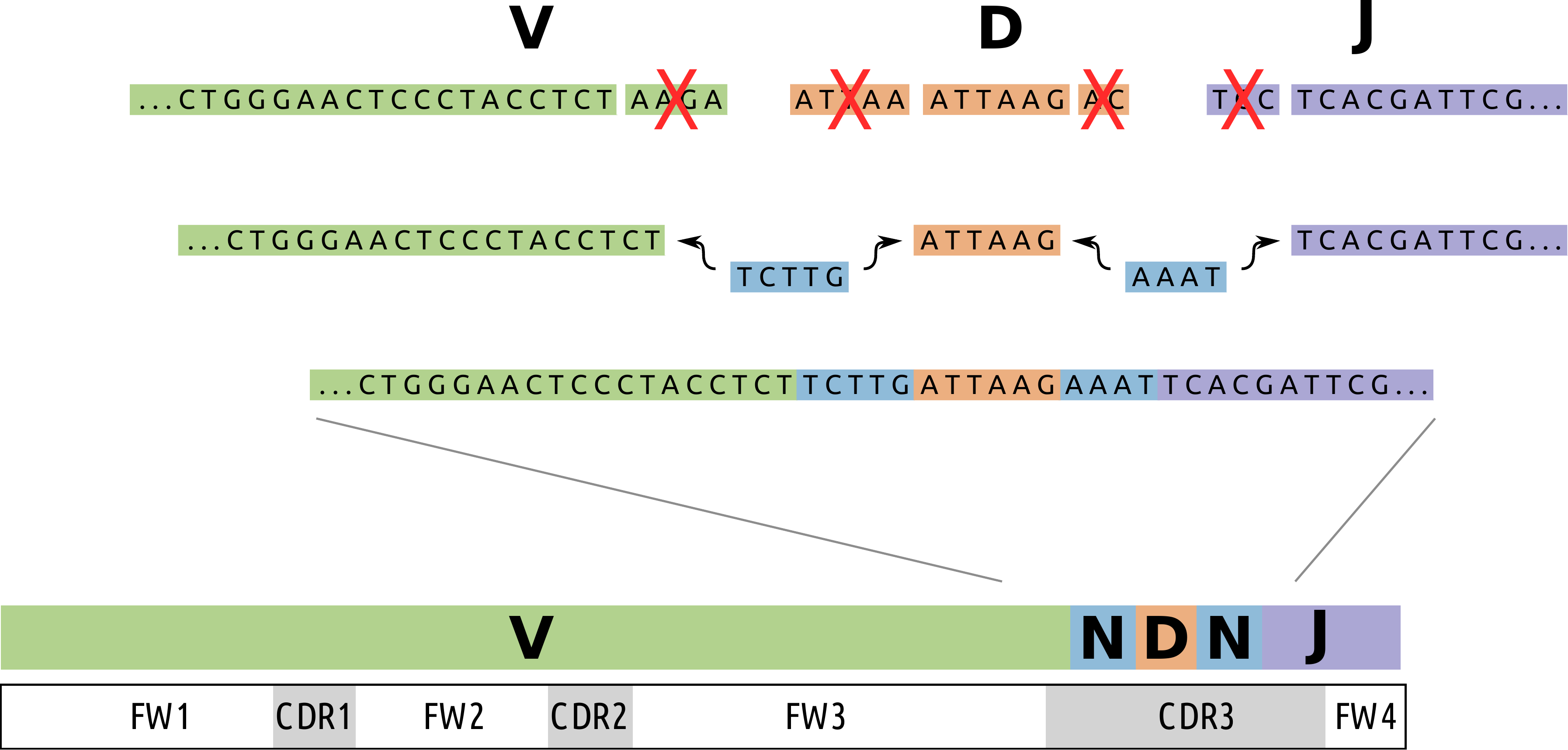}}
    \caption{\
      {\bf The \vdj\ recombination process}, in which individual V, D, and J genes are first randomly selected from a number of copies of each.
      These genes are then joined together via a process that deletes some randomly distributed number of nucleotides on their boundaries then joins them together with random ``non-templated'' nucleotides in the N-region (blue).
      The specificity of an antibody is to a large extent determined by the region defined by the heavy chain recombination site, referred to as the third complementarity determining region (CDR3).
    }
    \label{FIGaRecoEvent}
  \end{figure}
}

\newcommand{\FIGaFewHmmStates}{\
  \begin{figure}
    \begin{center}
      \forarxiv{
        \includegraphics[width=4.5in]{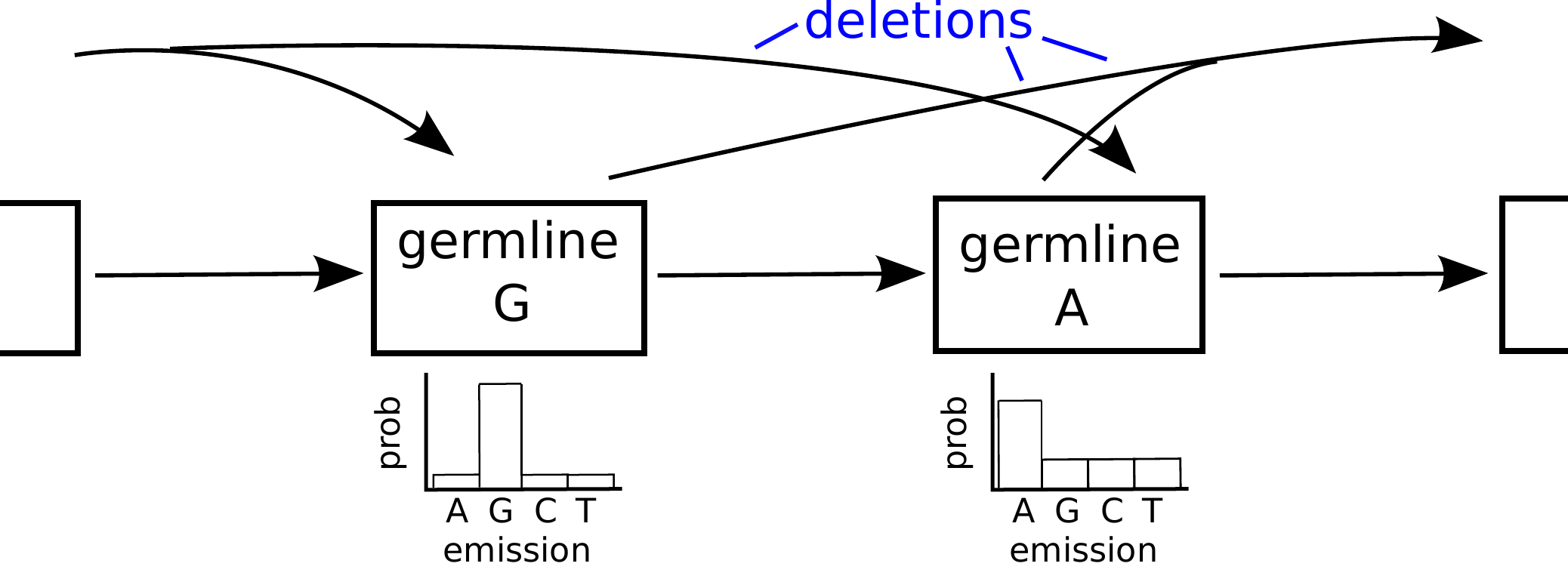}
      }
    \end{center}
    \caption{\
      {\bf A few states in the internal region of an HMM.}
      On the left is a state representing a position in a gene with a germline G, which usually emits a G, but sometimes emits other bases (i.e. mutates).
      If this state is near enough to the start or end of the gene, there will likely be transitions from the initial state, or to the end state (upper arrows).
      On the other hand, if it is in the middle of a gene the path is more likely to simply traverse the states in order (straight horizontal arrows).
    }
    \label{FIGaFewHmmStates}
  \end{figure}
}

\newcommand{\FIGhmmTopology}{\
  \begin{figure}
    \begin{center}
      \forarxiv{
        \includegraphics[width=4.5in]{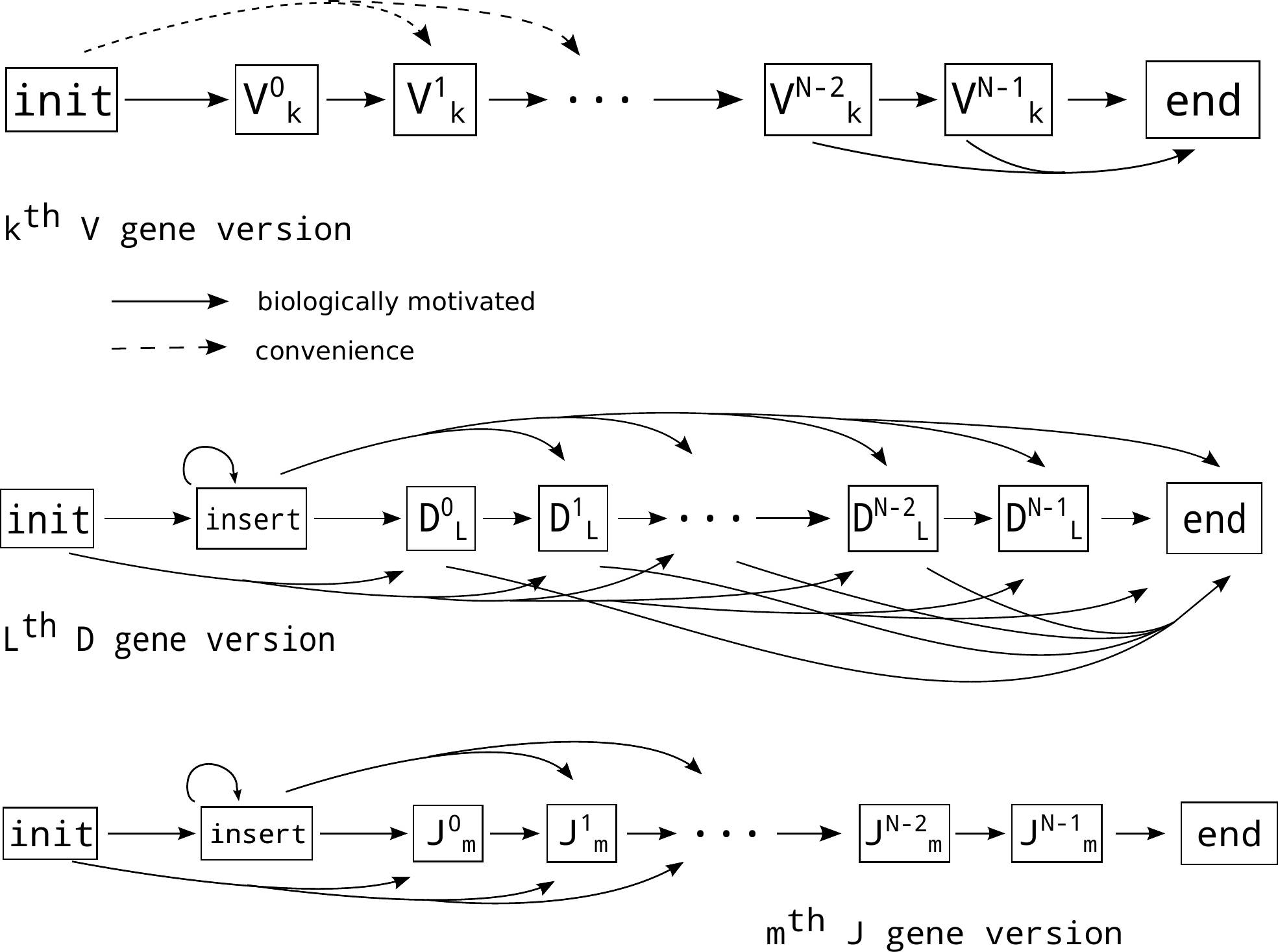}
      }
    \end{center}
    \caption{\
      {\bf Overall topology of the HMMs in the V, D, and J segments.}
      Inserts are shown as a single state for clarity, but are replaced by four states in the actual HMM (Figure~\ref{FIGfourInsertStates}).
      Note that we include 5' V and 3' D exonuclease deletions as a convenience (dashed lines) to account for varying read lengths.
    }
    \label{FIGhmmTopology}
  \end{figure}
}

\newcommand{\FIGfourInsertStates}{\
  \begin{figure}
    \begin{center}
      \forarxiv{
        \includegraphics[width=2.5in]{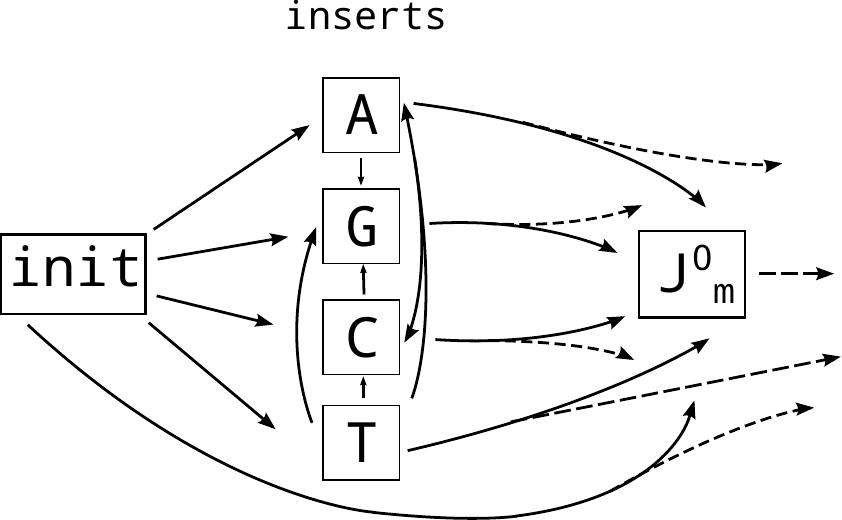}
      }
    \end{center}
    \caption{\
      {\bf HMM N-region topology.}
      By using four states instead of one, this gives improved discrimination for pairs or tuples of sequences because it does not ignore mutation information within N-regions.
      Self-transitions for insert states ommitted for clarity.
    }
    \label{FIGfourInsertStates}
  \end{figure}
}

\newcommand{\FIGmeanVariance}{\
  \begin{figure}
    \begin{center}
      \forarxiv{
        \includegraphics[width=\textwidth]{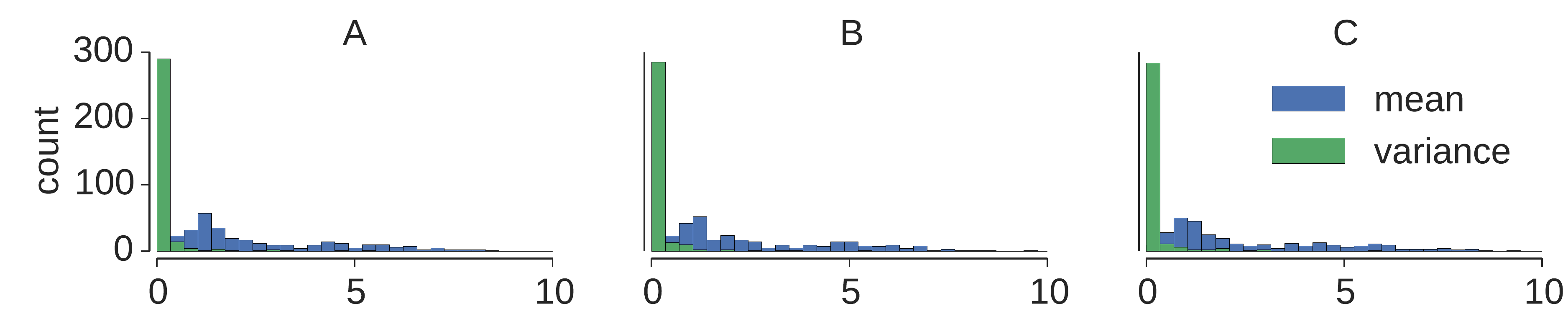}
      }
    \end{center}
    \caption{\
      {\bf Mean and variance of inferred parameters.}
      The across-subset mean and variance of inferred parameter values for each human in the Adaptive data set across 10 disjoint subsets of the data.
    }
    \label{FIGmeanVariance}
  \end{figure}
}

\newcommand{\FIGhumanVsHumanMutation}{\
  \begin{figure}
    \begin{center}
      \forarxiv{
        \includegraphics[width=\textwidth]{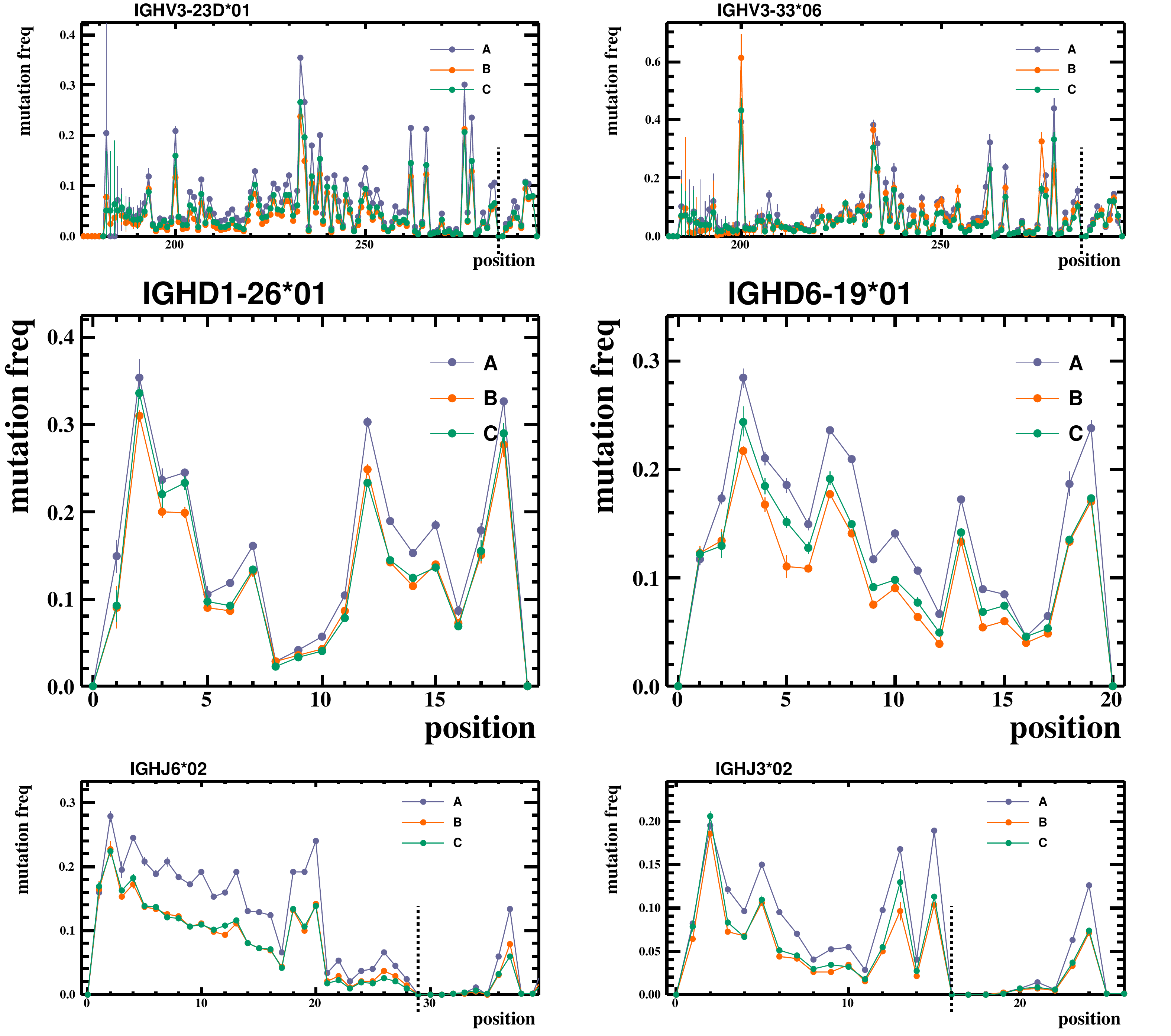}
      }
    \end{center}
    \caption{\
      {\bf Mutation frequency versus position.}
      Typical observed mutation frequencies for two V, two D, and two J alleles on the three humans (A, B, and C) in the Adaptive data set.
      The x axis is the zero-indexed position along the IMGT germline allele.
      Mutation frequencies are seen to be highly position-dependent.
      While the structure of these mutation distributions appears similar between humans, the overall level of mutation varies.
      The first base of the conserved cysteine and tryptophan codons (i.e. the CDR3 boundaries) are indicated with black vertical dashed lines.
      In the complete set of plots (which are publicly available as described in the text), mutation frequencies are highly variable across sites with a pattern that is similar between humans.
    }
    \label{FIGhumanVsHumanMutation}
  \end{figure}
}

\newcommand{\FIGhumanVsHumanDeletion}{\
  \begin{figure}
    \begin{center}
      \forarxiv{
        \includegraphics[width=0.8\textwidth]{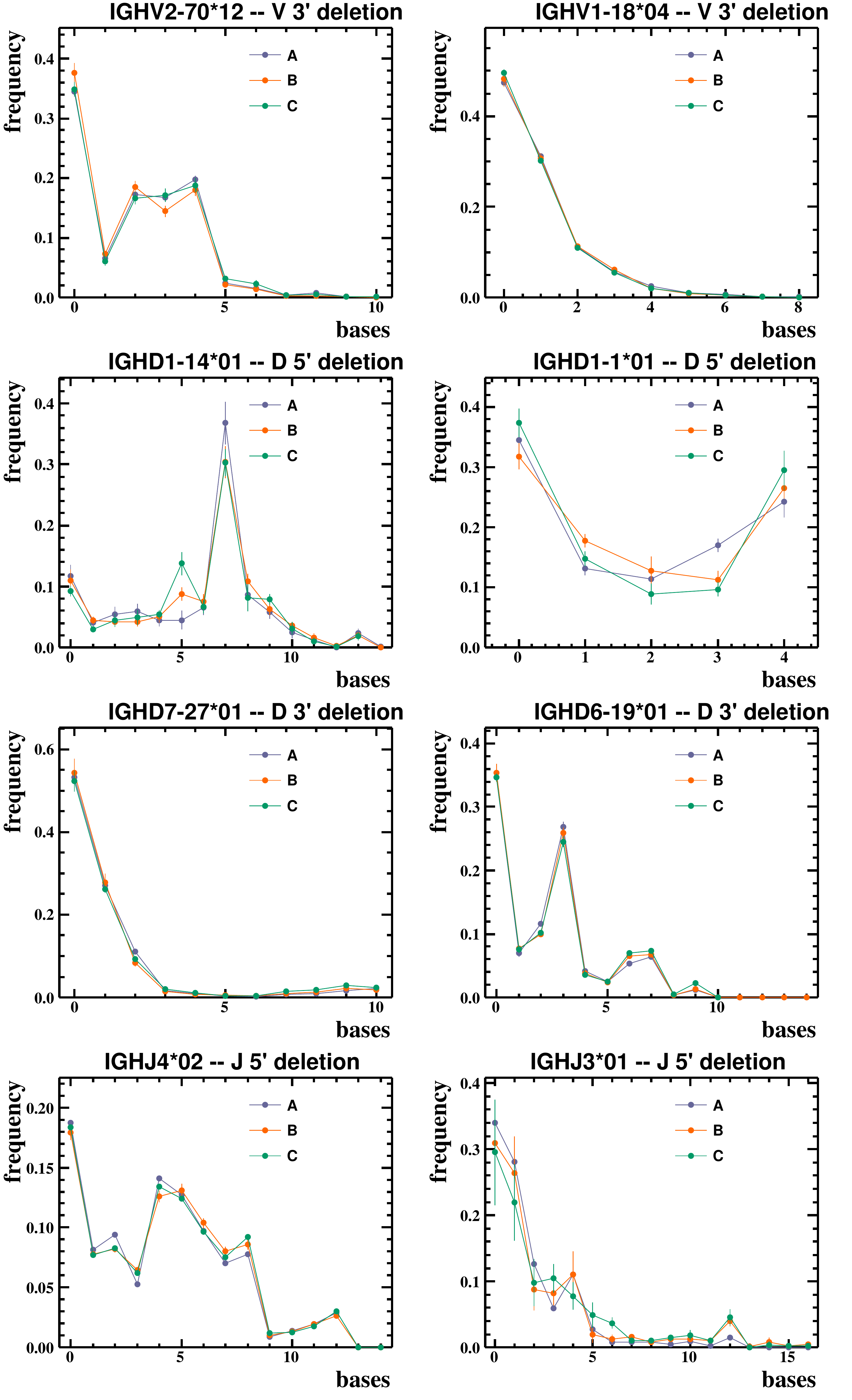}
      }
    \end{center}
    \caption{\
      {\bf Deletion length.}
      Observed exonuclease deletion length frequencies for two V, four D, and two J alleles on the three humans (A, B, and C) in the Adaptive data set.
      These alleles were chosen to be representative of the various shapes taken by the empirical distributions.
      In the complete set of plots (which are publicly available as described in the text), per-allele distributions are frequently multi-modal and appear similar between humans.
    }
    \label{FIGhumanVsHumanDeletion}
  \end{figure}
}

\newcommand{\FIGhumanVsHumanInsertionLengths}{\
  \begin{figure}
    \begin{center}
      \forarxiv{
        \includegraphics[width=\textwidth]{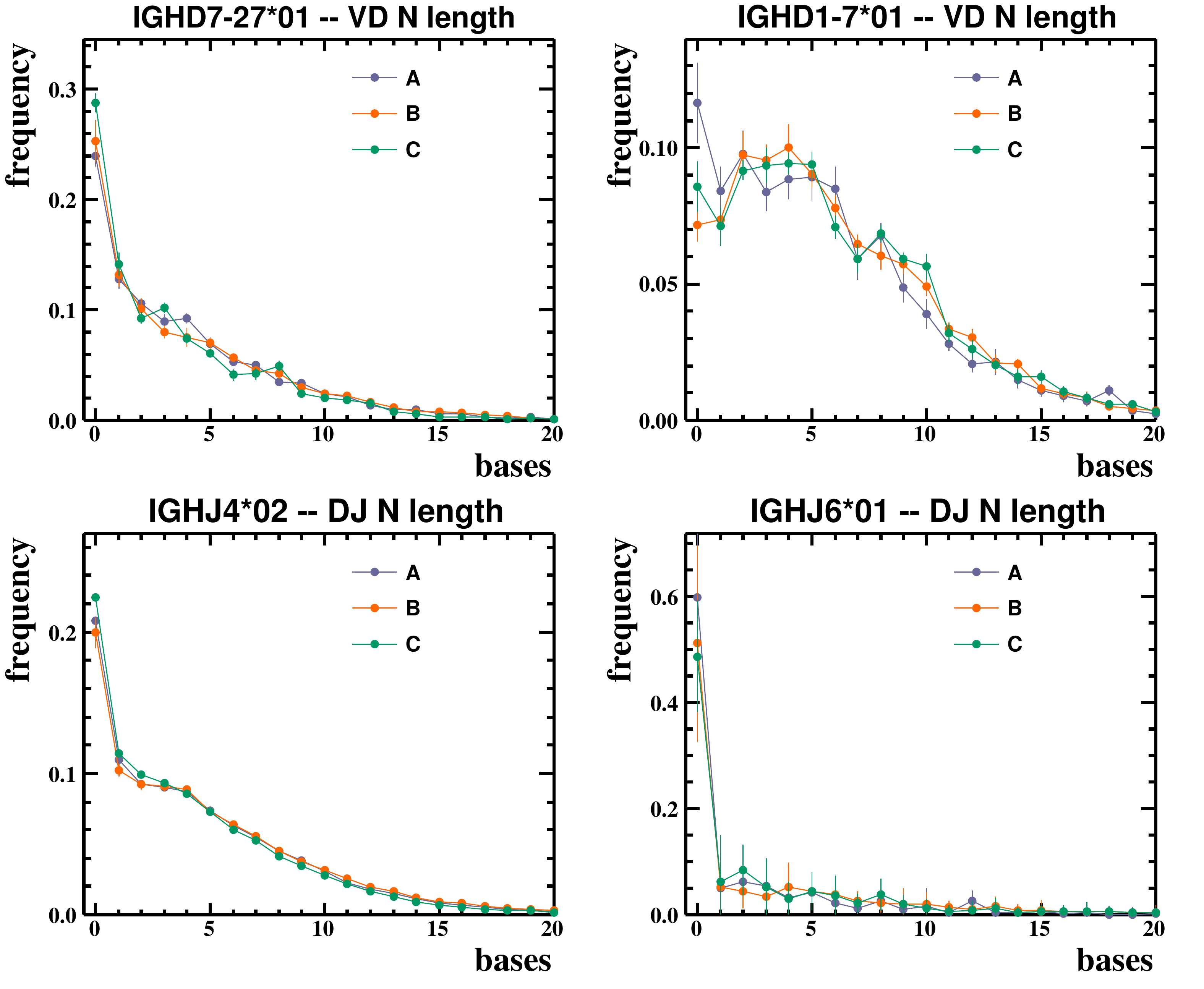}
      }
    \end{center}
    \caption{\
      {\bf N-region lengths.}
      Typical observed N-region lengths at the VD and DJ boundaries for two D and two J alleles.
      In the complete set of plots (which are publicly available as described in the text), distributions have a similar shape and the per-allele plots appear similar between humans.
    }
    \label{FIGhumanVsHumanInsertionLengths}
  \end{figure}
}

\newcommand{\FIGhumanVsHumanMutationVollmers}{\
  \begin{figure}
    \begin{center}
      \forarxiv{
        \includegraphics[width=\textwidth]{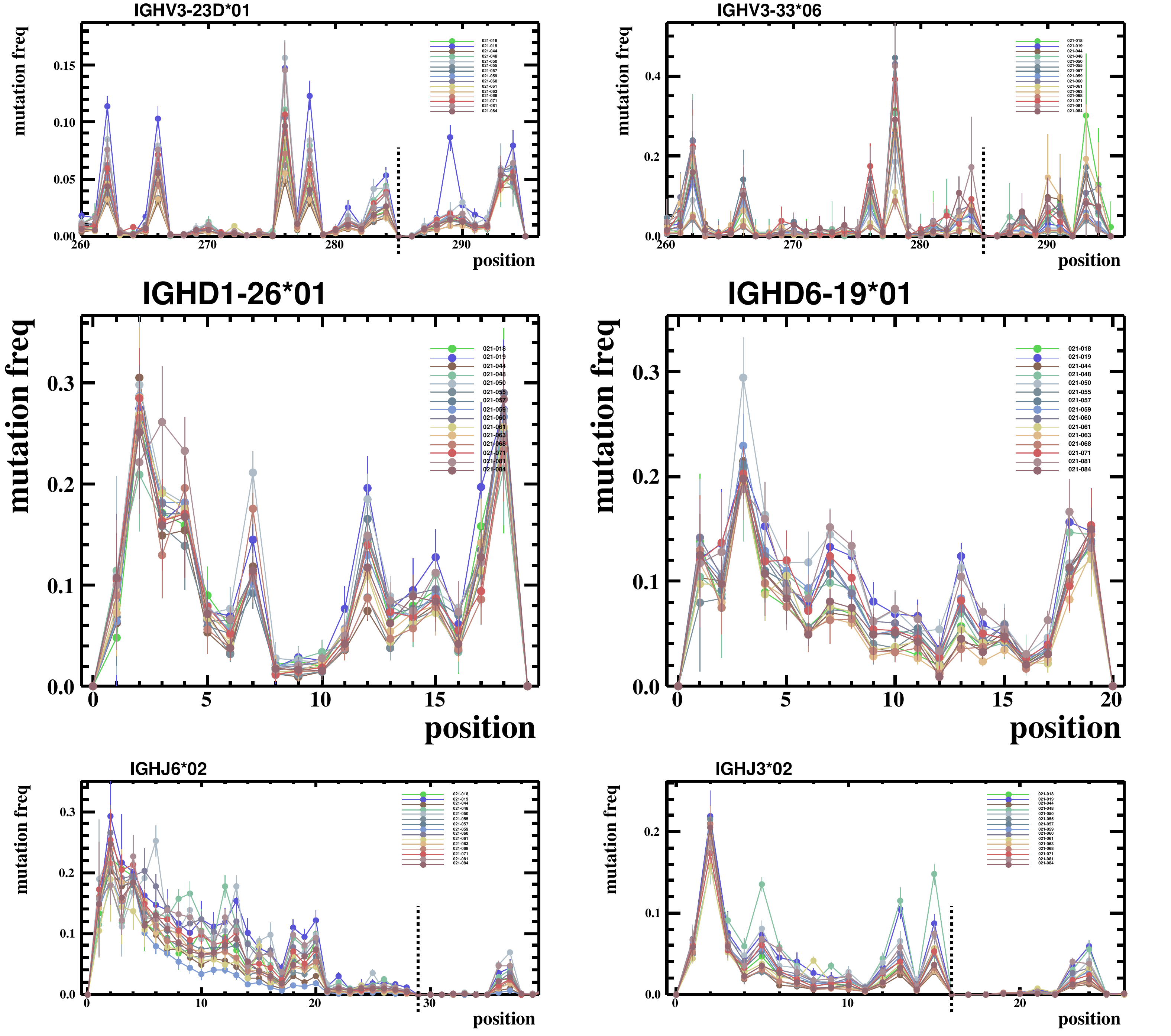}
      }
    \end{center}
    \caption{\
      {\bf Mutation frequency versus position.}
      Typical observed mutation frequencies for two V, two D, and two J alleles on the Vollmers data set.
      The first base of the conserved cysteine and tryptophan codons (i.e. the CDR3 boundaries) are indicated with black vertical dashed lines.
      The large uncertainties at the 5' end of V and 3' end of J reflect that our reads very rarely extend into these regions.
      Here and elsewhere we use the standard naming convention for alleles of germline genes: for example ``IGHV3-33*06'' means the 6th allele of the 33rd gene in the V3 gene family for the heavy chain.
    }
    \label{FIGhumanVsHumanMutationVollmers}
  \end{figure}
}

\newcommand{\FIGhumanVsHumanDeletionVollmers}{\
  \begin{figure}[ht]
    \begin{center}
      \forarxiv{
        \includegraphics[width=0.8\textwidth]{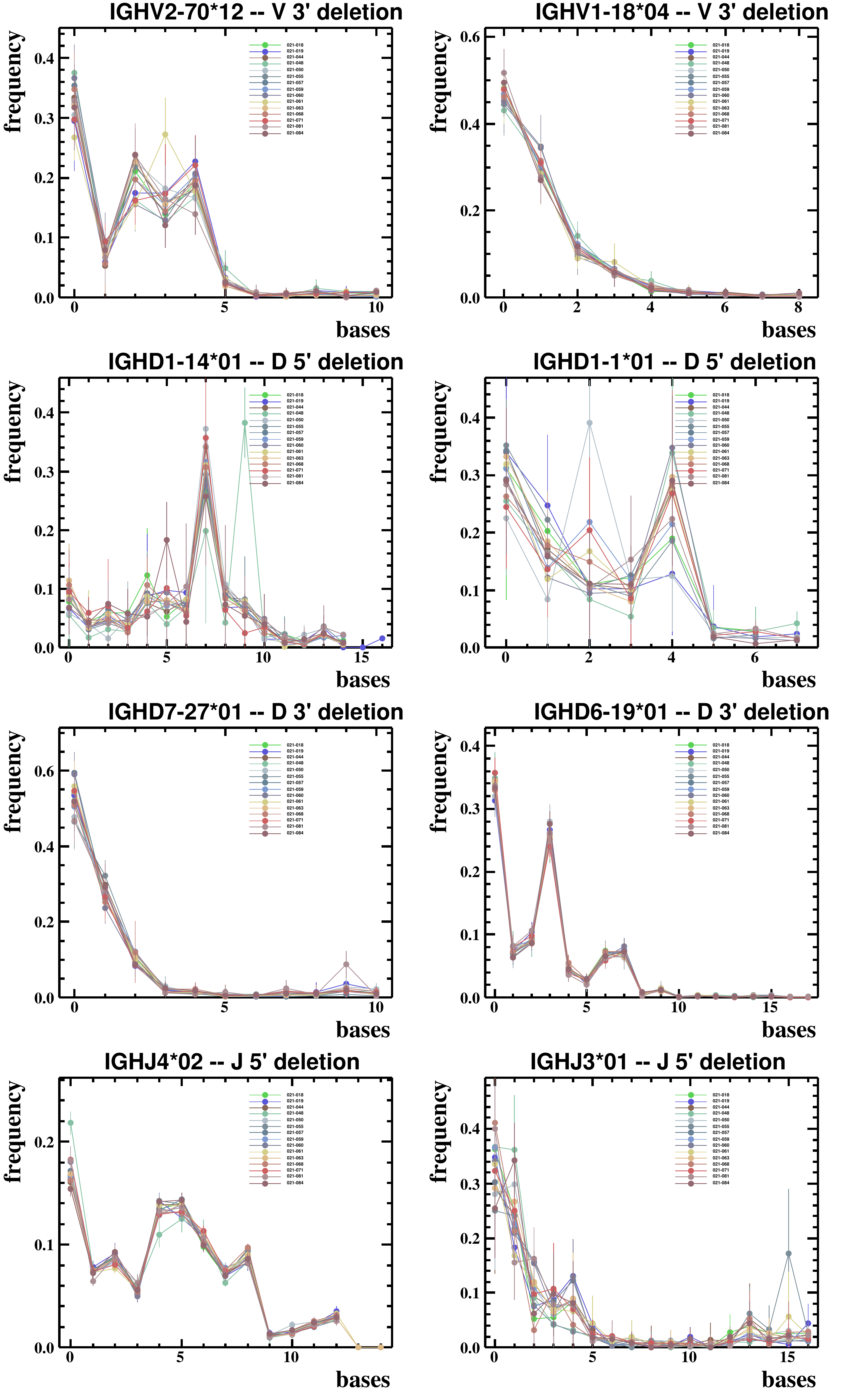}
      }
    \end{center}
    \caption{\
      {\bf Deletion lengths.}
      Typical observed exonuclease deletion length frequencies for two V, four D, and two J alleles on the Vollmers data set.
    }
    \label{FIGhumanVsHumanDeletionVollmers}
  \end{figure}
}

\newcommand{\FIGhumanVsHumanInsertionLengthsVollmers}{\
  \begin{figure}
    \begin{center}
      \forarxiv{
        \includegraphics[width=\textwidth]{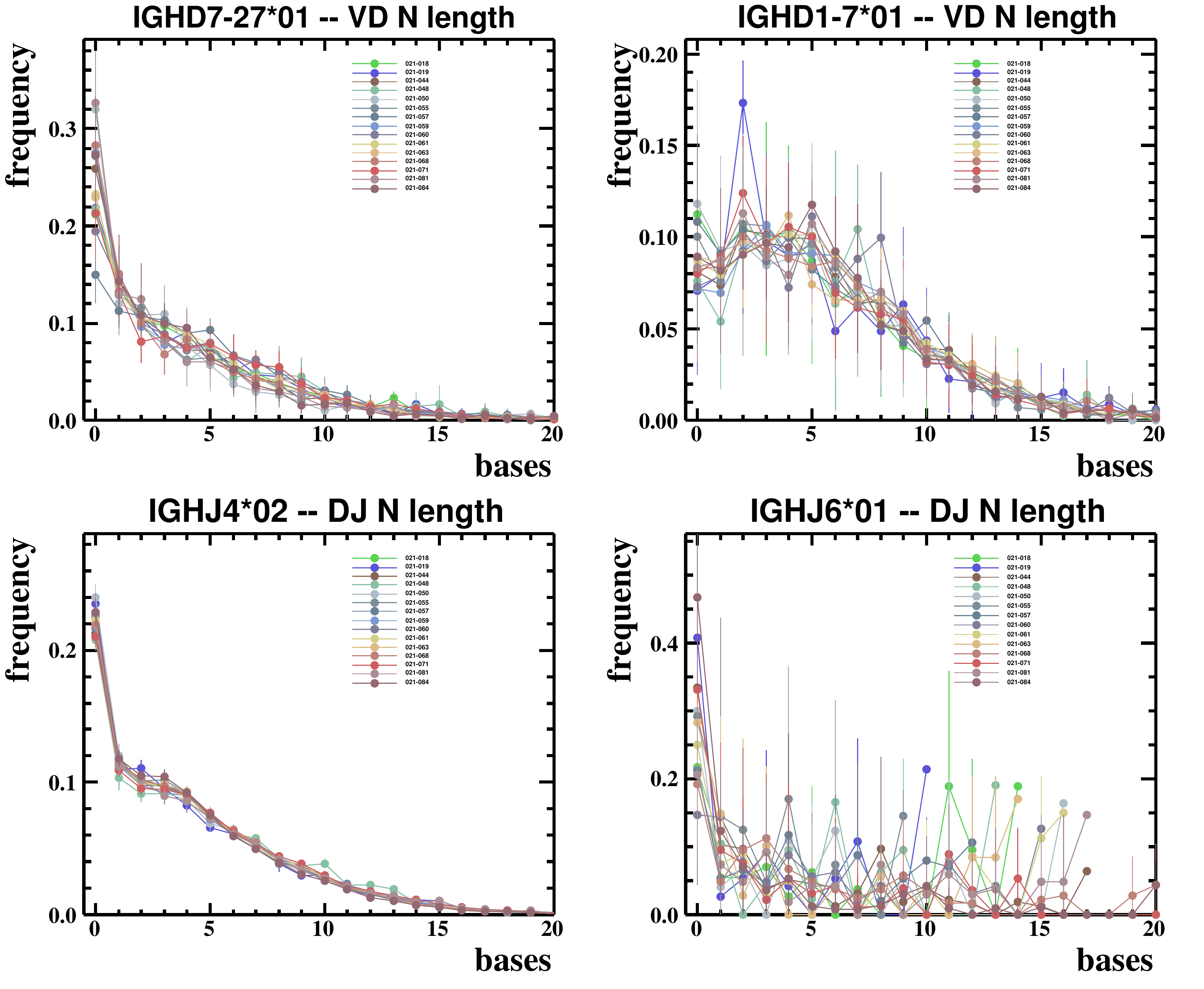}
      }
    \end{center}
    \caption{\
      {\bf N-region lengths.}
      Typical observed N-region lengths at the VD and DJ boundaries for two D and two J alleles.
    }
    \label{FIGhumanVsHumanInsertionLengthsVollmers}
  \end{figure}
}

\newcommand{\FIGmeanVarianceVollmers}{\
  \begin{figure}
    \begin{center}
      \forarxiv{
        \includegraphics[width=\textwidth]{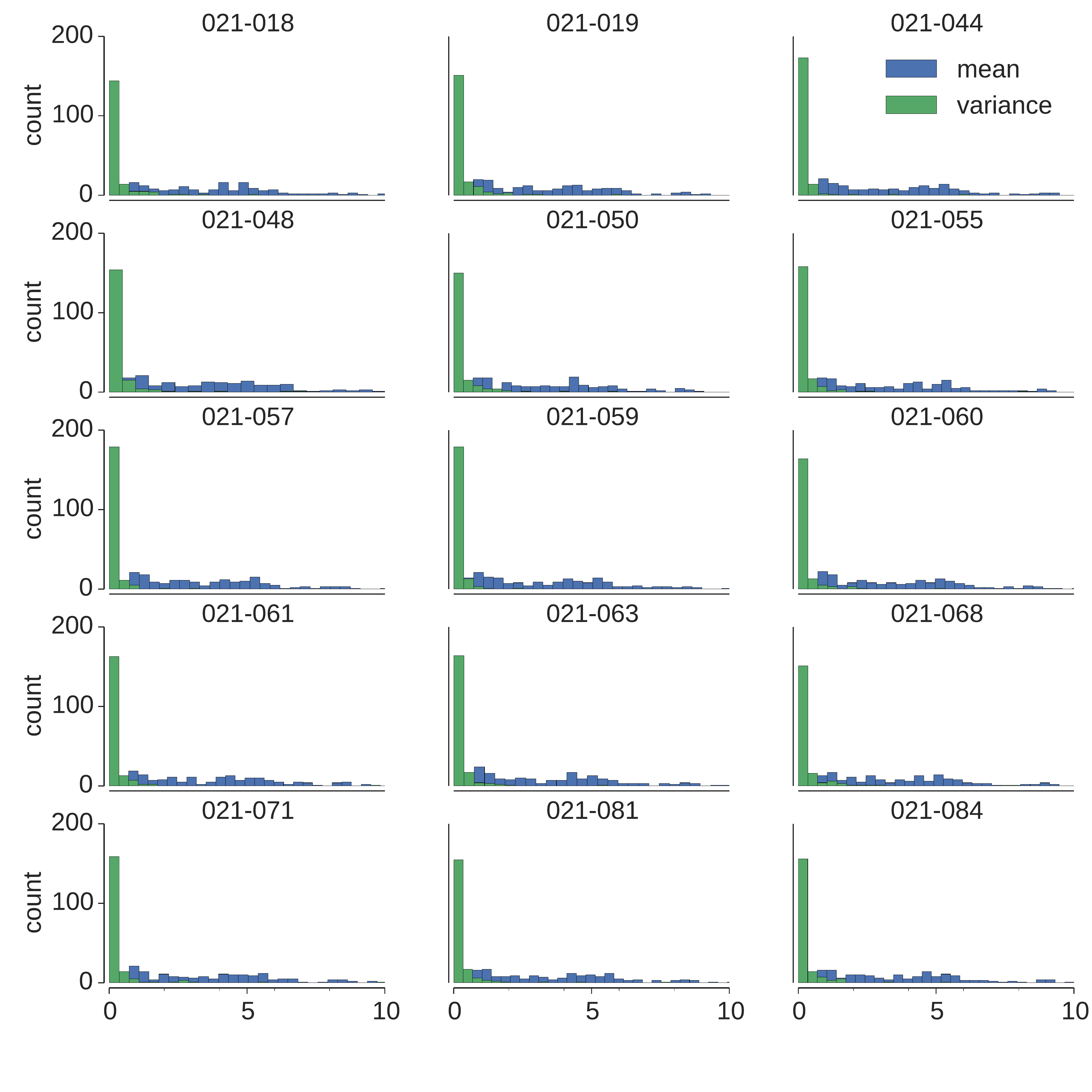}
      }
    \end{center}
    \caption{\
      {\bf Mean and variance of inferred parameters.}
      The across-subset mean and variance of inferred parameter values for each human in the Vollmers data set across 10 disjoint subsets of the data.
      See the caption to Figure~\ref{FIGmeanVariance} and the corresponding text for more details.
    }
    \label{FIGmeanVarianceVollmers}
  \end{figure}
}

\newcommand{\FIGdataVsSimuDeletionInsertion}{\
  \begin{figure}
    \begin{center}
      \forarxiv{
        \includegraphics[width=\textwidth]{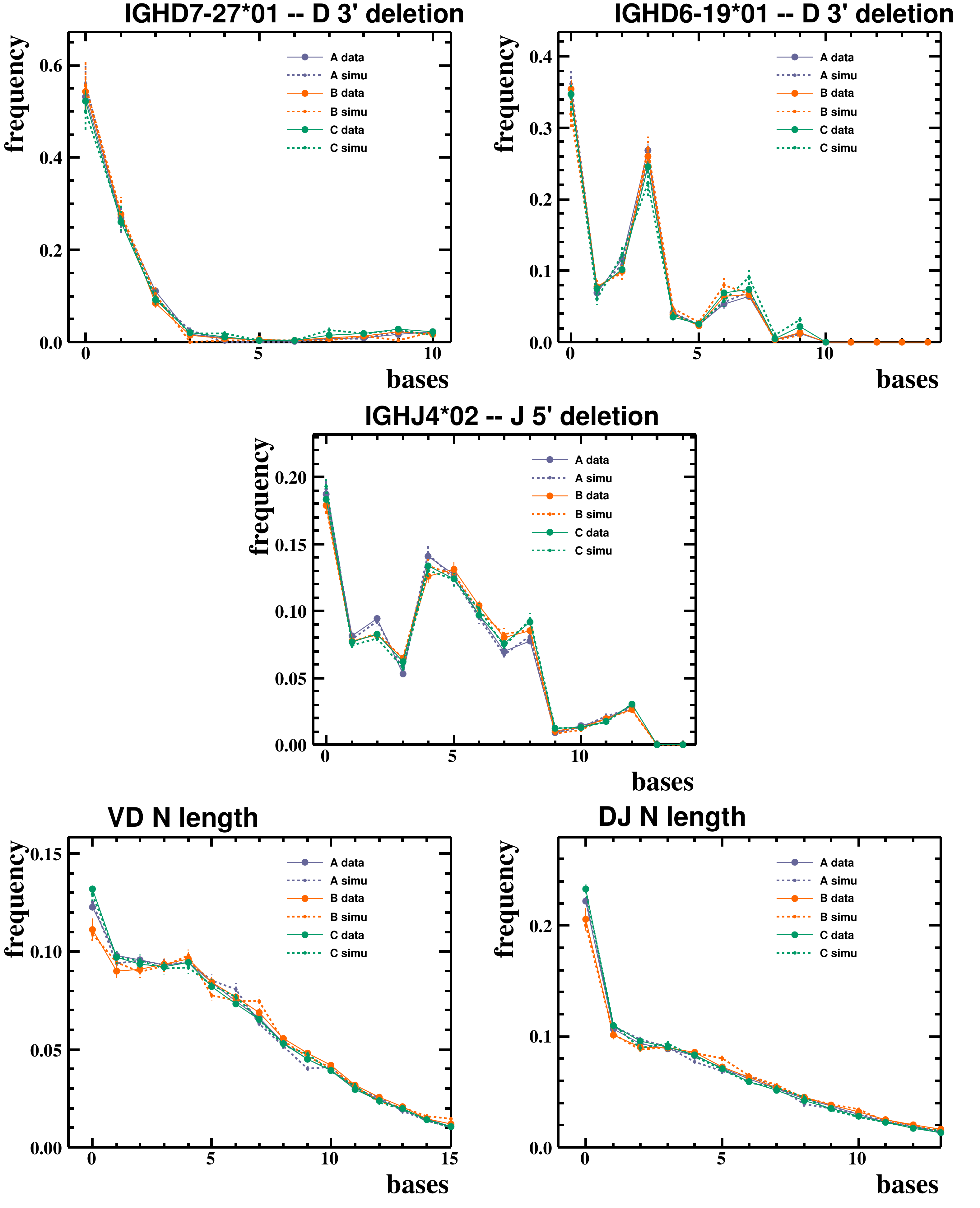}
      }
    \end{center}
    \caption{\
      {\bf Deletions and N-region lengths for data and simulation} for three different humans.
    }
    \label{FIGdataVsSimuDeletionInsertion}
  \end{figure}
}

\newcommand{\FIGdataVsSimuOverallMutation}{\
  \begin{figure}
    \begin{center}
      \forarxiv{
        \includegraphics[width=\textwidth]{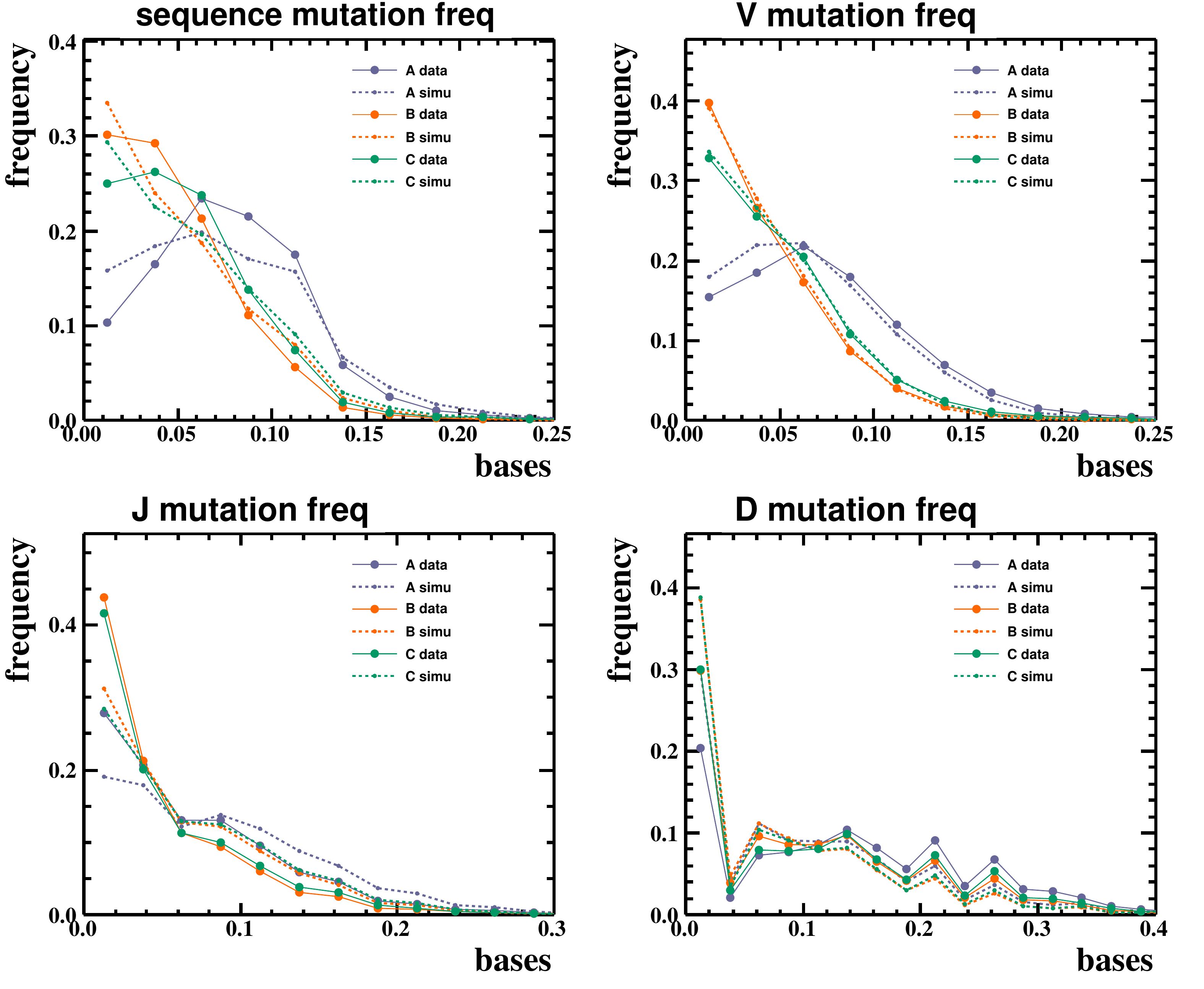}
      }
    \end{center}
    \caption{\
      {\bf Mutation frequencies for data and simulation} for three different humans over the full reads, and for the V, D, and J segments individually.
    }
    \label{FIGdataVsSimuOverallMutation}
  \end{figure}
}

\newcommand{\FIGdataVsSimuMutationPerPosition}{\
  \begin{figure}
    \begin{center}
      \forarxiv{
        \includegraphics[width=\textwidth]{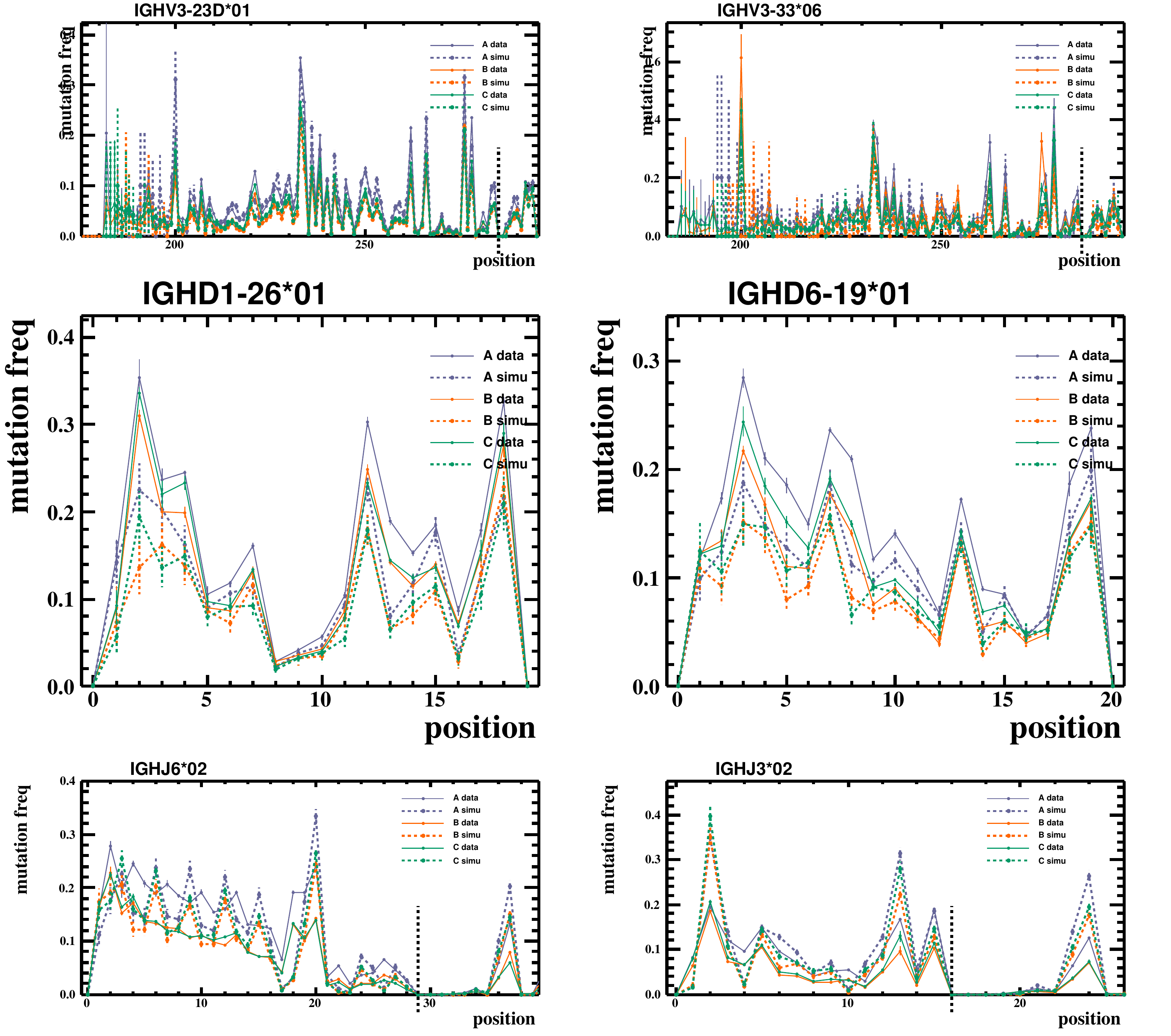}
      }
    \end{center}
    \caption{\
      {\bf Per-position mutation frequencies for data and simulation} for typical alleles in the V, D, and J segments.
    }
    \label{FIGdataVsSimuMutationPerPosition}
  \end{figure}
}

\newcommand{\TABLEfractionCorrectGenes}{\
  {\small
    \begin{table}[ht]
      \centering
      \caption{\bf Correct gene calls for all public BCR annotation methods}
      \begin{tabular}{|l|c|c|c|c|}
        \hline
             method      & V correct  & D correct & J correct   &  failed  \\ \hline
        \partis\ ($k=5$) & 0.9968 $\pm$.0006 &  0.828 $\pm$.004  &  0.9983 $\pm$.0005  &  0 $\pm$.00003  \\ \hline
        \partis\ ($k=1$) & 0.9955 $\pm$.0008 &  0.753 $\pm$.005  &  0.9766 $\pm$.0017  &  0 $\pm$.00003  \\ \hline
        \ighutil         & 0.9938 $\pm$.0009 &  0.714 $\pm$.005  &  0.961  $\pm$.002   &  0$\pm$.00003   \\ \hline
        \ihmmunealign    & 0.938  $\pm$.003  &  0.530 $\pm$.006  &  0.799  $\pm$.005   &  0.18$\pm$.004  \\ \hline
        \igblast         & 0.986  $\pm$.001  &  0.521 $\pm$.006  &  0.872  $\pm$.004   &  0.10$\pm$.003  \\ \hline
        \imgt            & 0.988  $\pm$.001  &  0.574 $\pm$.006  &  0.935  $\pm$.003   &  0.06$\pm$.003  \\ \hline
      \end{tabular}
      \caption*{\
        Fraction of genes correctly inferred up to allele, i.e. the number of sequences for which the specified method called the correct gene (regardless of allele) divided by the total number of sequences, for all publicly-available BCR annotation methods on a simulated sample of 30,000 sequences.
        For those methods which fail for some input sequences (usually due to poor D matches), we specify the fraction of sequences for which such failures occurred.
        \partis\ is shown both for single sequences ($k=1$), and for a multi-HMM ($k=5$), which performs simultaneous inference on five clonally related sequences.
        Values are shown $\pm$ the half-width of the binomial proportion 95\% confidence intervals using Jeffreys prior.
      }
      \label{TABLEfractionCorrectGenes}
  \end{table}}
}

\newcommand{\FIGgeneCorrectVsMutFreq}{\
  \begin{figure}
    \begin{center}
      \forarxiv{
        \includegraphics[width=\textwidth]{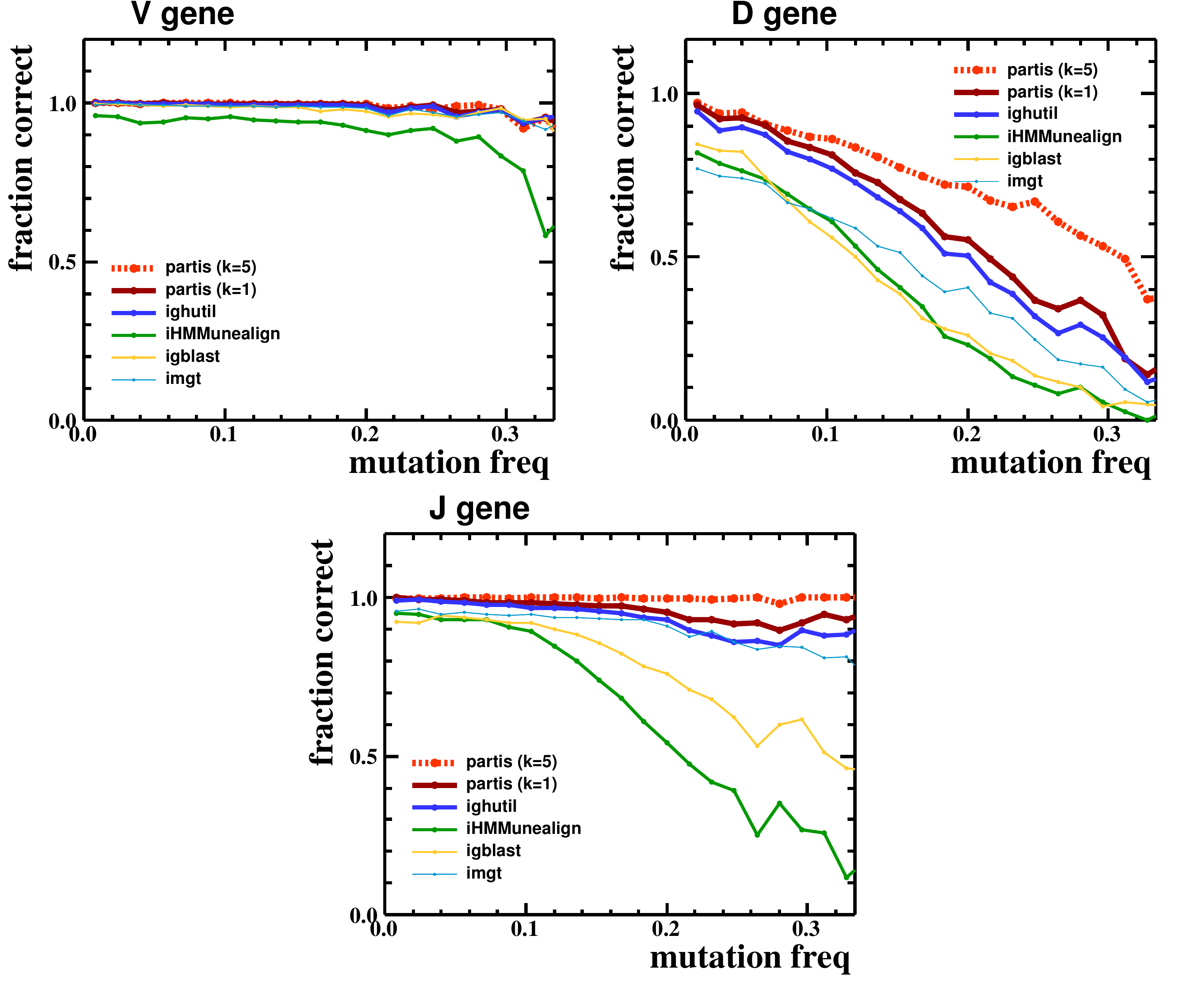}
      }
    \end{center}
    \caption{\
        {\bf Correct gene calls versus sequence mutation frequency}, i.e. the number of sequences for which the specified method called the correct gene (regardless of allele) divided by the total number of sequences, versus sequence mutation frequency, for publicly-available BCR annotation methods on a simulated sample of 30,000 sequences.
        \partis\ is shown both for single sequences ($k=1$), and for a multi-HMM ($k=5$), which performs simultaneous inference on five clonally related sequences.
    }
    \label{FIGgeneCorrectVsMutFreq}
  \end{figure}
}

\newcommand{\FIGhttn}{\
  \begin{figure}
    \begin{center}
      \forarxiv{
        \includegraphics[width=0.45 \textwidth]{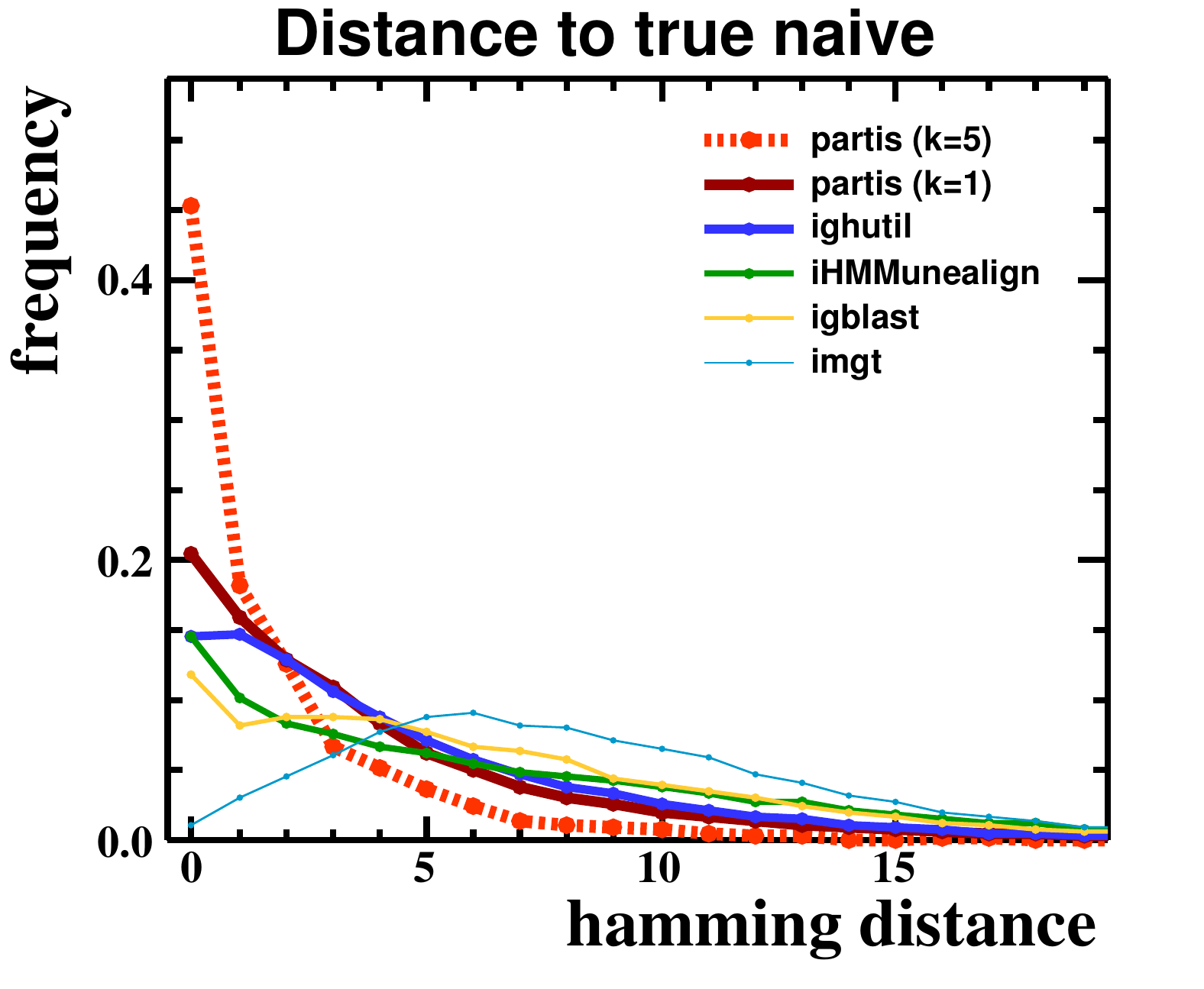}
      }
    \end{center}
    \caption{\
      {\bf Hamming distance between inferred and true naive sequences}, for all available BCR annotation methods on a simulated sample of 30,000 sequences.
      \partis\ is shown both for single sequences ($k=1$), and for a multi-HMM ($k=5$), which performs simultaneous inference on five clonally related sequences.
    }
    \label{FIGhttn}
  \end{figure}
}

\newcommand{\FIGotherMetrics}{\
  \begin{figure}
    \begin{center}
      \forarxiv{
        \includegraphics[width=\textwidth]{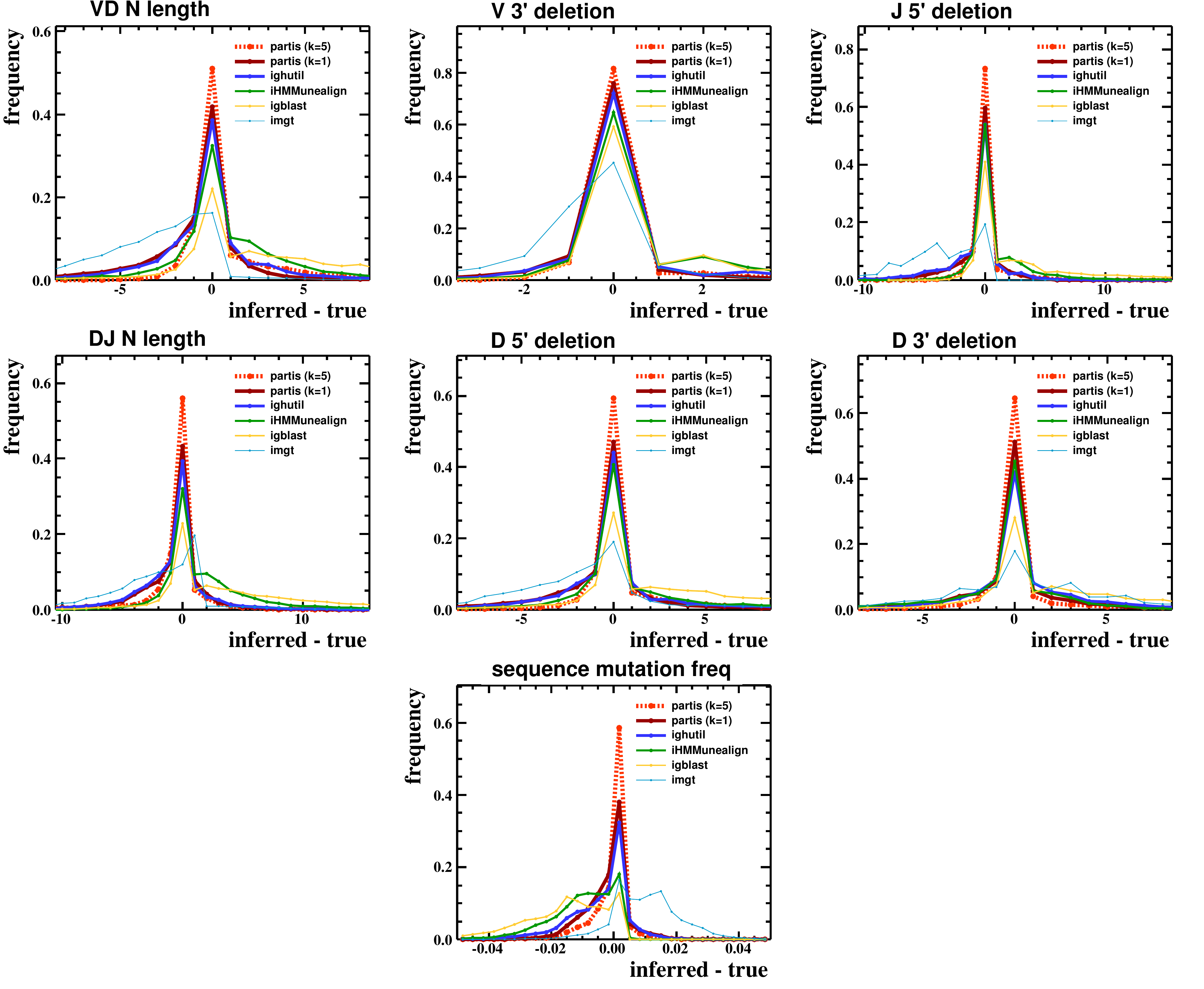}
      }
    \end{center}
    \caption{\
      {\bf True versus inferred parameters for all methods}. The difference between inferred and true values for exonuclease deletion lengths, N-region lengths, and mutation frequency for all available BCR annotation methods on a simulated sample of 30,000 sequences; more tightly peaked around zero is better.
      \partis\ is shown both for single sequences ($k=1$) and for a multi-HMM ($k=5$), which performs simultaneous inference on five clonally related sequences.
    }
    \label{FIGotherMetrics}
  \end{figure}
}

\newcommand{\FIGhttnBySampleSize}{\
  \begin{figure}
    \begin{center}
      \forarxiv{
        \includegraphics[width=0.45 \textwidth]{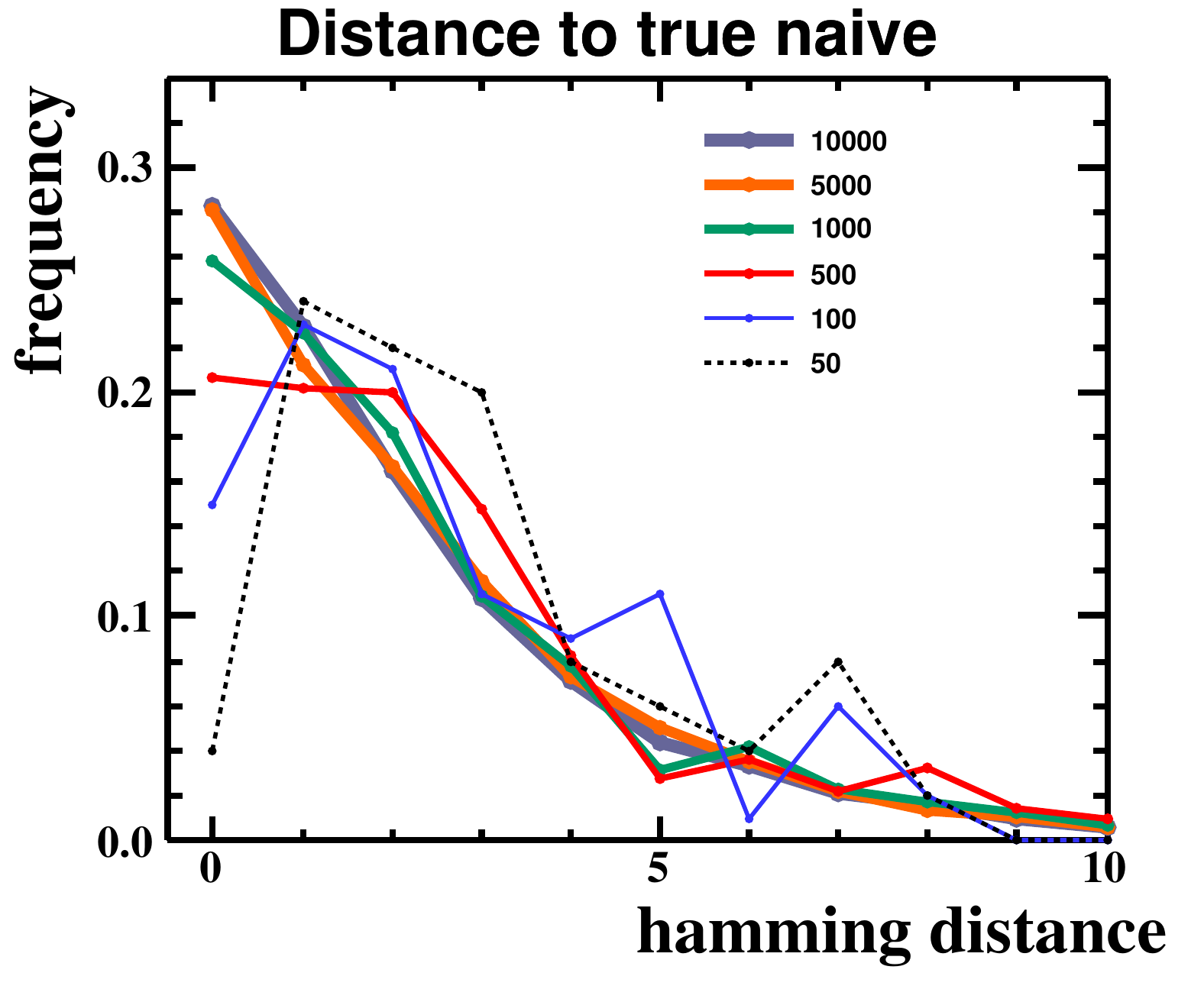}
      }
    \end{center}
    \caption{\
      {\bf Performance versus sample size.} \partis\ performance as measured by Hamming distance between inferred and true naive sequences when given 50 to 10,000 total sequences.
      When fewer sequences than a threshold are provided for a gene, \partis\ uses a tiered aggregation strategy (see Methods) to obtain enough sequences on which to do parameter estimation, which partly recovers performance as shown here.
    }
    \label{FIGhttnBySampleSize}
  \end{figure}
}

\begin{document}
\vspace*{0.35in}

\begin{flushleft}
{\Large
\textbf\newline{Consistency of VDJ rearrangement and substitution parameters enables accurate B cell receptor sequence annotation}
}
\newline
\\

Duncan Ralph\textsuperscript{1,\textcurrency},
Frederick A. Matsen IV\textsuperscript{1,*}
\\
\bigskip
\bf{1} Fred Hutchinson Cancer Research Center, Seattle, Washington, USA
\bigskip

\textsuperscript{\textcurrency} 1100 Fairview Ave N, Seattle, WA, 98109

\textsuperscript{*} matsen@fredhutch.org

\end{flushleft}

\section*{Abstract}
VDJ rearrangement and somatic hypermutation work together to produce antibody-coding B cell receptor (BCR) sequences for a remarkable diversity of antigens.
It is now possible to sequence these BCRs in high throughput; analysis of these sequences is bringing new insight into how antibodies develop, in particular for broadly-neutralizing antibodies against HIV and influenza.
A fundamental step in such sequence analysis is to annotate each base as coming from a specific one of the V, D, or J genes, or from an N-addition (a.k.a.\ non-templated insertion).
Previous work has used simple parametric distributions to model transitions from state to state in a hidden Markov model (HMM) of VDJ recombination, and assumed that mutations occur via the same process across sites.
However, codon frame and other effects have been observed to violate these parametric assumptions for such coding sequences, suggesting that a non-parametric approach to modeling the recombination process could be useful.
In our paper, we find that indeed large modern data sets suggest a model using parameter-rich per-allele categorical distributions for HMM transition probabilities and per-allele-per-position mutation probabilities, and that using such a model for inference leads to significantly improved results.
We present an accurate and efficient BCR sequence annotation software package using a novel HMM ``factorization'' strategy.
This package, called \partis\ (\url{https://github.com/psathyrella/partis/}), is built on a new general-purpose HMM compiler that can perform efficient inference given a simple text description of an HMM.

\section*{Author Summary}
The binding properties of antibodies are determined by the sequences of their corresponding B cell receptors (BCRs).
These BCR sequences are created in ``draft'' form by VDJ recombination, which randomly selects and deletes from the ends of V, D, and J genes, then joins them together with additional random nucleotides.
If they pass initial screening and bind an antigen, these sequences then undergo an evolutionary process of mutation and selection, ``revising'' the BCR to improve binding to its cognate antigen.
It has recently become possible to determine the BCR sequences resulting from this process in high throughput.
Although these sequences implicitly contain a wealth of information about both antigen exposure and the process by which humans learn to resist pathogens, this information can only be extracted using computer algorithms.
In this paper, we employ a computational and statistical approach to learn about the VDJ recombination process.
Using a large data set, we find consistent and detailed patterns in the parameters, such as amount of V gene exonuclease removal, for this process.
We can then use this parameter rich model to perform more accurate per-sequence attribution of each nucleotide to either a V, D, or J gene, or an N-addition (a.k.a.\ non-templated insertion).

\section*{Introduction}
The molecular sequences of B and T cell receptors (BCRs and TCRs) determine what antigens will be recognized by these lymphocytes, with B cells recognizing antigens via immunoglobulins \cite{Cooper2015-pk} and T cells recognizing antigenic peptides presented by the major histocompatibility complex \cite{Huppa2003-ks}.
Together, BCRs and TCRs are able to bind to a great diversity of antigens due to their sequence-level diversity.
Diversity in the receptor loci is generated first by the process of \vdj\ recombination, in which germline-encoded V, D, and J genes are randomly selected, the gene ends are trimmed some random amount, and then joined together with random non-templated insertions forming the N-region (Fig.~\ref{FIGaRecoEvent}) \cite{Hozumi1976-yo,Tonegawa1983-qz}.
BCR sequences diversify further through the Darwinian process of somatic hypermutation and antigen selection \cite{Weigert1970-rk,McKean1984-pn}.

\FIGaRecoEvent

In the 30-odd years since the discovery of \vdj\ recombination by Susumu Tonegawa \cite{Hozumi1976-yo}, molecular sequencing has been widely applied to BCRs and TCRs to further our understanding of this process.
Recently, high throughput sequencing has given a remarkable perspective on the forces determining the immunoglobulin repertoire \cite{Boyd2009-ci,Weinstein2009-zy,Larimore2012-lo,Reddy2010-hz,Jackson2013-ad,Georgiou2014-sh}.
In addition to advancing basic science understanding, such sequencing is being applied to learn how antibodies develop against antigens of medical interest, such as through influenza vaccination \cite{Vollmers2013-vh}.
There have also been spectacular advances using this technology to understand the ontogeny of HIV broadly neutralizing antibodies, with the still-elusive goal of eliciting them with an HIV vaccine \cite{Sok2013-td,Zhu2013-em,Gao2014-ls,Kepler2014-pg}.

A fundamental step in the analysis of such a sequencing data set is to reconstruct the origin of each nucleotide in each sequence: whether it came from an N-addition or from a germline V, D, or J gene, and if so, which one and where.
Even if a complete collection of alleles (gene variants between individuals) for the germline V, D, and J genes were available, this problem would be challenging because exonuclease deletion obscures the boundaries between N-regions and germline V, D, and J gene sequences.
The BCR case is made more difficult by the processes of somatic hypermutation and clonal selection: if a BCR sequence does not match a germline gene in an area adjoining exonuclease removal, it may have come from a mutated germline position or an N-addition (non-templated insertion).
We will call this general problem of describing the source of each nucleotide in a BCR (or TCR) sequence the BCR (or TCR) ``annotation problem''.
Here we will focus on the more challenging BCR variant of the problem.

One approach is to leverage general-purpose tools for doing pairwise sequence search such as BLAST \cite{Altschul1990-oe} and Smith-Waterman local alignment \cite{Smith1981-al} for the annotation problem.
Adding BCR-specific aspects to these basic algorithms has resulted in a collection of useful tools such as \igblast\ \cite{Ye2013-ei} and the online annotation tool on the \imgt\ \cite{Lefranc2009-iu} website.

However, BCR sequence formation is quite complex (reviewed in \cite{Jackson2013-ad}) and this complexity invites a modeling-based approach, specifically in the framework of hidden Markov models (HMMs).
HMMs for sequence analysis consist of a directed graph on ``hidden state'' nodes with defined start and end states, with each node potentially ``emitting'' a nucleotide base or amino acid residue \cite{Durbin1998-uq,Eddy2004-jm}.
In the BCR case, the hidden states represent either (gene, nucleotide position) pairs or N-region nucleotides, and the emission probabilities incorporate the probability of somatic hypermutation at that base.
The HMM approach to BCR annotation has been elegantly implemented first in \soda\ \cite{Volpe2005-uk}, then \ihmmunealign\ \cite{Gaeta2007-mz}, and then \sodatwo\ \cite{Munshaw2010-mj}.
The transition probabilities for these previous HMM methods were modeled parametrically: specifically, they used the negative binomial distribution as first used in \cite{Jackson2004-zd}, and the emission probabilities come from the same mutation process across positions (even if the process is context-dependent)~\cite{Gaeta2007-mz}.

High throughput sequencing has become commonplace in the time since these HMMs were designed; this provides both a challenge and an opportunity for such model-based approaches.
It is a challenge because millions of distinct sequences are now available from a single sample of B cells, so methods must be efficient.
On the other hand, it is an opportunity because such large data sets offer the opportunity to develop and fit models with much more detail.

We hypothesized that large data sets would reveal reproducible fine-scale details in the probabilistic rearrangement process that could be used for improved inference.
The reproducibility of such details on a per-gene level is suggested by two papers from the same group: first, analogous results in T cells \cite{Murugan2012-ue}, and, more recently, similar results for B cells (independent to that presented here) \cite{Elhanati2015-ru}.
For annotation via HMMs, researchers have previously used probability distributions such as the negative binomial \cite{Jackson2004-zd} to model exonuclease deletion lengths, and modeled the propensity for somatic hypermutation based on sequence context \cite{Gaeta2007-mz}.
However, BCR sequences are protein-coding, and thus there are constraints on N-region and exonuclease deletion lengths that come from sequence frame; these types of constraints cannot be expressed by unimodal probability distributions with few parameters.
For example, the post-selection D gene is preferentially (though not exclusively) used in a specific frame \cite{Larimore2012-lo,Benichou2013-ej}, which in some cases is simply due to a stop codon being present in the alternate frames \cite{Jackson2013-ad}.
This suggests that a parameter-rich approach could be useful: instead of parametric distributions with a few parameters, using a large data set we could fit a per-allele categorical distribution for N-region and exonuclease deletion lengths, in which every outcome has its own probability.
Rather than modeling the process of somatic hypermutation with germline nucleotide context, we could simply infer the per-position per-allele mutation frequency and use this as the mutation probability.

In order to meet the challenge and opportunity of large-scale data, we have built \partis: a fast, flexible, and open source HMM framework to analyze BCR sequences.
We started by writing an efficient new HMM compiler, called \ham, which enables inference on an arbitrary HMM specified via a simple text file rather than having to write special-purpose computer code.
We then developed an HMM ``factorization'' strategy, which along with extensive caching and optimizations in the overall model topology, results in an order of magnitude faster execution than previous HMM implementations for BCR/TCR sequences.
This work, along with parallelization, means that data sets of tens of millions of unique sequences \cite{DeKosky2015-ut,billion}, the largest available today, are comfortably within the capabilities of \partis\ running on standard research hardware.

We find that HMM inference using this approach to fitting HMM parameter distributions for heavy chain BCRs outperforms previous approaches for the annotation problem \cite{Volpe2005-uk,Gaeta2007-mz,Munshaw2010-mj}.
The parameters for the non-parametric transition and emission probabilities are fit ``on the fly'' for each individual data set, using a tiered aggregation strategy to scale the level of model detail to the amount of data.
The accuracy of \partis\ is further increased by using multiple sequences from the same rearrangement event (which differ due to somatic hypermutation) using pair-HMMs and their generalization to more than two emission sequences, which we will call ``multi-HMMs''.
We also leverage the full power of HMMs by not only calculating the best (known as Viterbi) annotations of a sequence according to that HMM, but also computing forward probabilities for single and multi-HMMs.
These forward probabilities allow one to integrate out uncertainty in quantities that are not of interest, retaining more accurate estimates (with uncertainty) of those parameters that are of interest.
The \partis\ software implementation has been engineered to be extensible and maintainable: it is open source and includes continuous integration and reproducible validation using the Docker software container system \cite{Boettiger2014-mm}.

\section*{Results}

\subsection*{Empirical distributions of VDJ recombination process parameters deviate reproducibly from simple distributions}

We find that the probability distributions of rearrangement parameters for heavy chain BCR sequences deviate reproducibly from any commonly used distribution.
For example, our inferred exonuclease deletion length distributions take on a variety of shapes (Fig.~\ref{FIGhumanVsHumanDeletion}), which vary from gene to gene, and allele to allele, but appear consistent across humans (see also Fig.~\ref{FIGhumanVsHumanDeletionVollmers}).
Mutation rates vary by an order of magnitude from position to position, creating a pattern which is unique to each allele but similar across humans for that allele (Fig.~\ref{FIGhumanVsHumanMutation} and~\ref{FIGhumanVsHumanMutationVollmers}).
The overall level of mutation, however, varies between humans in our data sets (as noted previously \cite{McCoy2014-vq}).
N-region lengths, on the other hand, vary much less across alleles, and are typically unimodal (Fig.~\ref{FIGhumanVsHumanInsertionLengths} and~\ref{FIGhumanVsHumanInsertionLengthsVollmers}).
The full collection of plots is available on Dryad at \url{http://datadryad.org/review?doi=doi:10.5061/dryad.149m8}.
In these plots the error bars are constructed using a bootstrapping procedure: we divide unique sequences in the data into ten subsets and plot the uncertainty as the standard deviation over these subsets (see Methods).

\FIGhumanVsHumanDeletion
\FIGhumanVsHumanMutation
\FIGhumanVsHumanInsertionLengths

By considering all of the parameter estimates together, we find that estimates are consistent between sequence subsets.
Indeed, we partitioned the set of unique sequences into 10 disjoint sets and estimated HMM parameters (transition and emission probabilities) independently for each subset to learn to what extent those parameters are dependent on details of the data set rather than underlying processes.
By plotting histograms of the means and variances for all of the HMM parameters, we find that the between-subset variance is consistently much smaller than the mean (Figs.~\ref{FIGmeanVariance} and~\ref{FIGmeanVarianceVollmers}).
That these parameter estimates are consistent between disjoint subsets of a collection of unique sequences, and to a lesser extent consistent between humans, indicates that they reflect biology rather than noise.
Although the suggestion of consistency between humans is interesting, our strategy of per-sample parameter fitting does not require any such assumption, and so we refrain from formalizing this suggestion of between-human consistency into a statistical statement.

\FIGmeanVariance

\subsection*{\ham}
We implemented a new HMM compiler, called \ham, because none of the existing compilers were suitable for our needs (see Methods for more details).
This program is able to perform classical, pair, and multi-HMM inference on any HMM within computational constraints given a text-file description of its topology.
This text file description is in YAML format (\url{http://yaml.org/}), which is easy to write both by hand and with existing libraries for all popular programming languages.

\TABLEhamPerformance
Although our initial motivations lay more in usability than optimization, \ham's general purpose C++ code is slightly faster and uses somewhat less memory than code generated by the well-known \hmmoc\ \cite{Lunter2007}, even ignoring the extra time for code generation and compilation in the latter (Table~\ref{TABLEhamPerformance}).
The \ham\ C++ code is open source (GPL v3 license), and is available at \url{https://github.com/psathyrella/ham/}.

\subsection*{\partis}
We then used \ham\ to build a platform called \partis\ to perform annotation of B cell receptor sequences (BCRs) using HMMs.
This platform is especially effective for modern large sequencing data sets because it can infer detailed parameters concerning the rearrangement process, and then perform annotation inference on each sequence in the set.
It is written in Python, is open source (GPL v3 license), and is available at \url{https://github.com/psathyrella/partis/}.
A Docker image with \partis\ installed is available at \url{https://registry.hub.docker.com/u/psathyrella/partis/}.

In order to find a maximum-likelihood annotation, \partis\ calculates the Viterbi path through the HMM, giving an annotation consisting of the germline genes used to make the BCR, the amount of junctional exonuclease deletion for each gene, and the size of junctional insertions (N-regions) between the trimmed germline genes.
It can also compute forward probabilities, which give the joint probability of sequence and annotation.
All of this information is output as text files in comma-separated format.

\subsection*{Benchmarking on simulated data}
We developed a simulation engine for B cell receptor sequences implemented in non-HMM-based code separate from that used for inference (see Methods).
This engine implicitly takes into account detailed dependencies between variables beyond what is possible in our inferential framework.
To create a ``true naive'' (i.e.\ unmutated ancestor) sequence it begins by sampling a single point from the joint empirical distribution over all rearrangement parameters observed in a data set:
such a draw specifies which V, D, and J genes led to the BCR, the amount by which they are trimmed, and the length of the N-regions.
Because the draw is from the joint empirical distribution, it reflects our best understanding of a ``true'' rearrangement event from the data set.
The simulator then generates a random sequence with corresponding rearrangement parameters (e.g. N-region length between V and J), and then simulates somatic hypermutation out to each leaf of a given tree using per-position mutation rates.
The test sample for which we show performance comparisons (Table~\ref{TABLEfractionCorrectGenes} and Figs.~\ref{FIGgeneCorrectVsMutFreq},~\ref{FIGhttn}, and~\ref{FIGotherMetrics}) may be found in the \partis\ GitHub repository at \url{http://git.io/Fxuk}.
This sample was generated with parameters from human A in the Adaptive data set, except that the mean mutation frequency was doubled in order to provide a more challenging inference test (the original mutation frequencies for the three humans can be found in Fig.~\ref{FIGdataVsSimuOverallMutation}).

\TABLEfractionCorrectGenes
\FIGgeneCorrectVsMutFreq
\FIGhttn
\FIGotherMetrics

We find that our strategy using flexible categorical distributions (i.e.\ general discrete distributions) outperforms previous BCR annotation packages on this simulated data.
Indeed, \partis\ has higher accuracy than other methods in terms of the simple fraction of genes that are correctly inferred (Table~\ref{TABLEfractionCorrectGenes}), a metric used in previous studies~\cite{Lott2014,Munshaw2010-mj}.
We note that this higher accuracy is true across a variety of mutation frequencies (Fig.~\ref{FIGgeneCorrectVsMutFreq}).
While intuitive, however, the fraction of correct genes ignores the difference between slightly incorrect and very incorrect inferences in terms of underlying sequence (see Methods for details).

A more detailed comparison via the Hamming distance between inferred and true naive sequences also shows that \partis\ outperforms previous HMM-based implementations (Fig.~\ref{FIGhttn}).
Neither \soda\ \cite{Volpe2005-uk} nor \sodatwo\ \cite{Munshaw2010-mj} were available from the web or directly from the author for comparison.
The \ighutil\ program \cite{ighutil, McCoy2014-vq} is a straightforward application of Smith-Waterman alignment \cite{Smith1981-al} to the annotation problem.

\partis\ also gave the most accurate parameter inferences (the narrowest distributions around the correct value in Fig.~\ref{FIGotherMetrics}) for N-region lengths, exonuclease deletion lengths, and mutation frequencies.

Additionally, we find that using multiple sequences from a single rearrangement event (differing only in somatic hypermutation) leads to more accurate inference than using one sequence at a time.
\ham's ability to simultaneously emit an arbitrary number of symbols allows us to take advantage of this in a way that previous implementations could not, providing an additional boost in accuracy.
This annotation with multi-HMMs is shown in the preceding Figures as ``\partis\ ($k=5$)'' because, as an example, we do inference on five sequences known to form a clonal family.

Although the benefits of our more-detailed model decrease with decreasing sample size, we find that our ``tiered aggregation'' strategy (see Methods) performs well with smaller numbers of sequences (Fig.~\ref{FIGhttnBySampleSize}).
Indeed, performance does not begin to degrade appreciably until sample size drops to around a hundred sequences.

\FIGhttnBySampleSize

We also find that \partis\ is computationally efficient, making it suitable for inference on large data sets.
For efficiency tests, we used samples of one and ten thousand simulated sequences.
The Smith-Waterman methods in \ighutil\ require about 0.03 seconds per sequence.
The HMM-based inference, alone, of \partis\ takes about 0.04 seconds per sequence; however \partis\ performs a preliminary Smith-Waterman step (see Methods), and typically also writes the HMM input files from the Smith-Waterman and initial HMM steps.
In total, then, \partis\ takes between 0.1 and 0.2 seconds per sequence, depending on sample size.
The other available HMM-based method, \ihmmunealign, takes about 5 seconds per sequence on both data sets.
We suspect that this gain in speed is from the HMM factorization scheme described in the Methods.
As a practical example, a sample with 100,000 sequences takes \partis\ about 20 minutes to annotate using ten processes on our code-development server (which has a hyperthreaded 8-core 2.9 GHz Xenon processor).
As a typical example on a research compute cluster, one might run one million sequences with 100 processes divided across some number of machines, which would also take about 20 minutes (job dispatch via \texttt{slurm} is easily via a \partis\ command line flag).
For centers without computing clusters, we estimate that analyzing a million unique sequences would cost \$5.25 using on-demand compute through Amazon Web Services.
We did not attempt speed comparison with web-based BCR analysis tools, although we note that as a group they cannot be scaled to the size of the task, and are subject to periods of high load and downtimes.

Shorter sequences, which do not include the entire V and/or J segments, are perfectly appropriate input for \partis.
The method currently requires only that there is at least one base from the V, D, and J, although of course longer sequences will lead to more accurate results.

\section*{Discussion}

We find substantial complexity in the details of the heavy chain BCR rearrangement process; these details are consistent between data subsets and appear consistent between humans.
This observation agrees with the corresponding observation for T cells \cite{Murugan2012-ue}, with work showing codon effects on D segment frame usage \cite{Larimore2012-lo,Benichou2013-ej}, and more recent work on B cells \cite{Elhanati2015-ru}.
This complexity suggests the use of parameter-rich categorical distributions for transition and emission probabilities in a hidden Markov model (HMM) for BCR annotation.
Indeed, we find that incorporating the peculiarities and richness of modern data sets enables more accurate annotation of BCR sequences.
By letting the data inform parameters such as per-position mutation frequencies, we side-step the need to perform sophisticated modeling of processes such as context-sensitive hypermutation.
The multi-HMM framework enables simultaneous annotation of a number of sequences in a clonal family, which results in increased accuracy because the effects of various mutations can be averaged out in a principled way.
The four insert states in our HMM topology also allow inference of unmutated ancestral sequences in N-regions.

We have extended the ideas of previous authors using HMMs for BCR sequence annotation~\cite{Gaeta2007-mz,Munshaw2010-mj,Volpe2005-uk} and previous work on HMM compilers~\cite{Lunter2007,Lott2014} to build \partis, a system to annotate BCR sequences.
We find that this system substantially outperforms previous methods when given simulated data.
The \partis\ package and its dependencies are open source, and validation can be run using a supplied Docker image.

When \partis\ is presented with a data set for the first time, it infers around one to ten thousand parameters for that data set; with this in mind, we would like to be very clear concerning the potential for over-fitting.
For any given application of \partis, these parameter estimates are intermediate steps for which we are not making a claim of generality, and are instead specific to that data set.
This situation is analogous to fitting the rate parameters of a mutation rate matrix when performing a maximum likelihood phylogenetic analysis.
In contrast, previous methods used a collection of sequences as a training set for parameter inference and applied HMMs with those general-purpose parameters to other sequences.
Thus it makes perfect sense that previous authors \cite{Munshaw2010-mj} test on simulated sequences made with a different set of parameters than the ones used for training.
However, because \partis\ fits parameters on the fly for each data set, it never makes out-of-sample predictions and thus does not ``train'' in this sense.
This is not to say that the approach would not benefit from some form of across-sample estimation of parameters.
If, for example, the sample taken is small and does not appropriately represent the true parameter distribution, then one would obtain too high of confidence in bad inferences from these incorrect parameter estimates.
Such situations could benefit from regularization via empirical Bayes, using such overall parameter estimates as across-human priors \cite{Robbins1956-cn}.
Such an approach will become increasingly viable as we become more familiar with the population-level distributions of the relevant parameters.

We have also presented a robust validation framework for \partis.
For effective validation we need a set of sequences which is both representative of the data on which we will apply the method, and for which we know the true rearrangement history.
Unfortunately real data, which is the most readily-available source of sequences which satisfy the first criterion, does not exist which also satisfies the second.
In~\cite{Gaeta2007-mz}, for instance, validation was performed on two data sets which were each known to consist of a single clone with unknown true annotation, and the authors used the fraction of sequences annotated to have the same germline alleles as a proxy for accuracy.
This approach cannot test whether the methods are inferring the correct allele, and more significantly, it tests on only two points in the entire, many-dimensional rearrangement space.
In other words, validation was performed on a sample of two rearrangements, and gives no information as to how each method might perform on every other possible combination of germline allele choices and exonuclease deletion boundaries.

We thus chose to use simulation, for which by construction one knows the correct answer, but which also requires proof that it accurately represents data (see Methods for details).
Our simulations cover a broad range of parameter space, and use parameter distributions inferred from data sets.
However, they do not explicitly model sequencing error, as the landscape of sequencing techniques is rapidly evolving, including techniques such as replicate immunosequencing \cite{DeWitt2014-sz, billion} and barcoding \cite{Vollmers2013-vh,He2014-fn}.
For the same reason, we have not attempted to add platform-specific emission probabilities to the HMM, and thus users employing \partis\ for the annotation of sequence data from a platform with a specific bias should be careful that this bias does not creep into the annotations.

The results presented here have all used the IMGT set of germline sequences.
We acknowledge the substantial discussion concerning this set's accuracy~\cite{Lee2006-ti,Gadala-Maria2015-uq}, but IMGT is still by far the most popular resource for germline information and tools.
We have thus designed \partis\ so it can switch to any other germline database using a simple command line flag, for instance as new information becomes available~\cite{Gadala-Maria2015-uq,Watson2014-op}.

We emphasize that the improved accuracy obtained using the multi-HMM framework in simulation assumed that we knew the collection of sequences form a clonal family.
On real data, of course, we do not know which sequences derive from the same rearrangement event, and in fact this is a challenging problem.
Our next step will be to use the HMM framework presented here to cluster sequences together by rearrangement event.

While we and others have found HMMs to be useful for the BCR annotation problem, they do have certain limitations.
The Markov assumption (that the current state is ignorant of all states except for the previous one) makes it difficult to propagate information across the HMM.
For example, levels of mutation are correlated between different segments of the BCR~\cite{McCoy2014-vq}; thus upon traversing the V segment of a query sequence we have information concerning the overall mutation rate in the rest of the sequence.
The Markov chain's conditional independence, however, makes it impossible to propagate this information to the D and J segments in a strict HMM framework.
Additionally, HMMs cannot account for palindromic N-additions~\cite{Saada2007-kf}, complex strand interaction events~\cite{Kepler1996-kd,Jackson2007-ue}, or the appearance of tandem D segments~\cite{Larimore2012-lo}.
Conditional Random Fields (reviewed in \cite{MAL-013}) could provide a way around some of these limitations; linear-chain conditional random fields enjoy many of the attractive computational properties of HMMs while allowing for more complex dependencies.
As a final note, the HMM does not at this time model insertion/deletion mutations within gene segments \cite{Kepler2014-jy}.
We instead look for such events during the preliminary Smith-Waterman alignment step, and if one is found the user may run the HMM both with and without the insertion/deletion in question.
Such events, though rare in the Adaptive data set~\cite{McCoy2014-vq}, could in principle be incorporated into an HMM with additional transitions between the states representing insertions and deletions within the HMM, although significant innovation would be needed in order to keep annotation inference tractable.

\section*{Methods}

\subsection*{Data sets used}
As our primary data set, we used the Illumina-sequenced heavy chain BCR sequence data generated from a single time point sample from each of three humans, as recently described in \cite{billion}.
This data set is distinguished by its size (30 million total unique sequences, split between three humans and naive versus memory compartments) and the experimental design, which used replicate wells to enable template count estimation and decrease error~\cite{DeWitt2014-sz}.
This data will be available at \url{http://adaptivebiotech.com/link/publicBCellResource} upon publication of \cite{DeWitt2014-sz}; we will call it the ``Adaptive'' data set.

We also show results on the data from \cite{Vollmers2013-vh}, which used Illumina paired-end sequencing of heavy chain BCRs to investigate memory B cell recall after vaccination in 15 humans.
This data set used unique barcodes to decrease error, a technique which was validated using a reproducibility experiment.
This data can be publicly accessed in GenBank (\url{http://www.ncbi.nlm.nih.gov/projects/gap/cgi-bin/study.cgi?study_id=phs000656.v1.p1}); we will call this the ``Vollmers'' data set.

\subsection*{\ham}
We considered writing specialized code to do HMM inference on B cell receptor sequences, as has been done by previous authors \cite{Volpe2005-uk,Gaeta2007-mz,Munshaw2010-mj}.
However, for maximum flexibility and usability, we were inspired by other work \cite{Lunter2007,Lott2014} to build an HMM ``compiler'' that can do HMM inference given a description of the HMM in a format that is simpler to interpret and modify than specialized inferential machinery.
We were greatly inspired by the \hmmoc\ \cite{Lunter2007} and \stochhmm\ \cite{Lott2014} tools, but neither of them fully satisfied our needs.
We found the XML input format in \hmmoc\ too complicated to script, and the paradigm of auto-generated C++ code to be extremely difficult to test and debug.
\stochhmm, meanwhile, had a custom configuration file format, lacked a pair-HMM implementation, and possessed an extraordinarily extensive but not entirely functional feature set.
However, \stochhmm's overall structure, and its basic idea of reading the HMM specification from a text file rather than generating code, were similar to what we desired, so we used it as a starting point for a complete rewrite.
(As an aside, we note that the excellent HMMER tool \cite{Eddy1998-gr} only implements profile HMMs, and thus is not appropriate for our needs.)

We were thus led to build a new tool, which we call \ham.
\ham\ uses the intuitive YAML text format, which is easy to read and write by hand for simple cases, while being equally simple to use for more complex models via existing programming libraries (such as for \href{http://pyyaml.org/}{python} and \href{https://code.google.com/p/yaml-cpp/}{C++}).
This makes it trivial to test many different HMM topologies, since such modifications only require editing a few lines in the text file.

In a simple comparison on the canonical ``occasionally dishonest casino'' example~\cite{Durbin1998-uq} we find it slightly faster and more memory-efficient than \hmmoc\ (Table~\ref{TABLEhamPerformance}).
To generate these results we ran the Viterbi algorithm using both programs on sequences of length one (resp.\ ten) million, and averaged CPU time and memory use over 300 (resp.\ 30) runs.
We note in passing that \hmmoc's XML for this example consists of 5961 characters, while \ham's YAML specifies the same information in only 440.

\ham\ can simultaneously emit onto an arbitrary number ($k$) of outputs, such as for a pair-HMM ($k=2$).
As previously described, this ability is useful for BCR ancestral inference.

We have also included an optimization we call ``chunk-caching'' which greatly speeds repetitive dynamic programing (DP) table calculations.
Any time the left-hand side of a sequence is the same as that of another sequence, the corresponding left hand side of their DP tables will be identical.
We have thus included the ability to cache and recall previous DP tables in order to avoid unnecessary recalculation in such cases.

\paragraph*{HMM architecture}
We follow previous work \cite{Volpe2005-uk,Gaeta2007-mz,Munshaw2010-mj} by representing each germline base in each V, D, and J allele as an HMM state (Fig.~\ref{FIGaFewHmmStates}).
N-additions are represented as a separate class of states, and 5' (3') exonuclease deletions as transitions to or from germline states which are not at the start (end) of the gene.
All of these states can be combined to create a single HMM for the entire \vdj\ rearrangement process.
While it can be useful to think of individually calculating the probability of each such path, in practice one traverses the dynamic programming (DP) table using the recursion relations found in~\cite{Durbin1998-uq}.

\FIGaFewHmmStates

\paragraph*{Factorization}
All inferential operations on the HMM described in the previous section can be made faster by a process of ``factorization'', which performs inference on a collection of smaller HMMs but returns exactly the same results as a monolithic HMM.
Each of these smaller HMMs has the topology obtained by extracting it from the monolithic HMM, resulting in the topologies shown in Fig.~\ref{FIGhmmTopology}.
In order to motivate the procedure, we observe that 1) once a path hits a state corresponding to a given allele, it cannot transition to any state corresponding to a different allele for that segment (for example, a path in allele $D_1$ cannot transition to $D_2$); and 2) the only external information needed to calculate the DP table on a single allele's HMM for a given query sequence is where in the query sequence to start and where to end.

\FIGhmmTopology

In order to describe the factorization process in more detail, we will use the notation of \cite{Durbin1998-uq} referring to a single HMM: $e_{s}(b)$ represents the probability of emitting $b$ in state $s$, and $a_{s s'}$ is the transition probability from state $s$ to $s'$.
We also use $\pi$ for a path through the set of hidden states, and $p_{i}(x, \pi)$ as an abbreviation for $e_{\pi_{i}}(x_{i}) a_{\pi_{i-1} \pi_{i}}$: the contribution of position $x$ to the forward probability.

One can write the total probability of a sequence $x$ under a given HMM, i.e. the forward probability, as a sum over paths $\pi$
\begin{equation}
  \label{eq:forward-prob}
  P(x) = \sum_{\pi} P(x, \pi).
\end{equation}
(For the Viterbi algorithm we would replace the summation with $\argmax$.)
As written, this corresponds to a single, monolithic HMM, and we are implicitly summing over all paths through all alleles in the V, D, and J segments.
In order to factorize this monolithic probability, we first write the probability of each path in more detail as a product over each position $i$ in the sequence of length $L$:
\begin{equation*}
  P(x, \pi) := \prod_{i=1}^{L} e_{\pi_{i}}(x_{i}) a_{\pi_{i-1} \pi_{i}} = \prod_{i=1}^{L} p_{i}(x, \pi).
\end{equation*}
Here $\pi_{i}$ is the state at position $i$ in path $\pi$, and $x_{i}$ is the symbol at position $i$ in the query sequence.
Thus $e_{\pi_{i}}(x_{i})$ is the probability of emitting $x_i$ from state $\pi_i$, and $a_{\pi_{i-1} \pi_{i}}$ is the transition probability from the state at $i-1$ to the state at $i$.

We now subdivide this product for the path $\pi$ into three factors for the V, D, and J segments; this divides the sequence correspondingly into three sections of length $l_{V}$, $l_{D}$, and $L-l_{V}-l_{D}$.
We will use $P_{R}(x, \pi, l_{V}, l_{D})$ to denote the forward probability of the path $\pi$ through a single segment $R \in [V, D, J]$.
\begin{eqnarray*}
  P(x, \pi) = \prod_{i=1}^{l_{V}} p_{i}(x, \pi)
              \prod_{j=l_{V}+1}^{l_{V}+l_{D}}  p_{j}(x, \pi)
              \prod_{k=l_{V}+l_{D}+1}^{L} p_{k}(x, \pi) \\
              = P_{V}(x, \pi, l_{V}, l_{D}) P_{D}(x, \pi, l_{V}, l_{D}) P_{J}(x, \pi, l_{V}, l_{D}) \\
              = \prod_{R = V, D, J} P_{R}(x, \pi, l_{V}, l_{D}).
\end{eqnarray*}
We then use this factorization to rearrange the order of operations in the sum over paths \eqref{eq:forward-prob}:

\begin{align*}
  \sum_{\pi} P(x, \pi)
  & = \sum_{\pi} \prod_{R = V, D, J} P_{R}(x, \pi, l_{V}, l_{D}) \\
  & = \sum_{l_{V}, l_{D}} \prod_{R = V, D, J} \ \sum_{g \in R} \ q_{g} \sum_{\pi \in g} P_{R}(x, \pi, l_{V}, l_{D}),
\end{align*}
where $q_{g}$ is the probability of choosing each allele $g$.
Here we have isolated the computationally expensive sum over all states $\sum_{\pi\in g}$ as far to the right as possible, so that we sum only over the paths within each allele, and combine all alleles afterward.
Factorization thus reduces a single HMM with a state for each germline position in each of a few hundred alleles, to a number of HMMs, each with states only for a single allele.
Since at each step in the forward (and Viterbi) recursion relations we sum over all possible previous states for each current state, computing time scales roughly as the square of the number of states in each HMM.
By reducing the maximum number of states per HMM by several orders of magnitude, factorization thus substantially decreases memory and computation time.

N-additions change this picture very little: we simply add them to the left side of each D and J HMM (they could as easily go on the right side of V and D).
The resulting HMM topologies are shown in Fig.~\ref{FIGhmmTopology}.
In this version, we show a single insert state, with emission probabilities corresponding to the empirical N-region nucleotide distribution from data (compare Fig.~\ref{FIGfourInsertStates}).

\FIGfourInsertStates

As described above, the HMM does not model insertion/deletion mutations within gene segments at this time.
Their inclusion would impose a significant computational burden: rather than a linear number of transitions there would be a quadratic number, and adding deletion self-transitions would require exploring the complete collection of potential deletions at every site in order to get a valid forward probability.
Noting that the Smith-Waterman step provides only slightly less accurate annotation than the HMM, we instead provide the option of filtering out potential insertion/deletion mutations during the initial Smith-Waterman step.
This allows the user to make a subjective decision, which we believe is appropriate in the face of the currently incomplete germline sequence databases, combined with many germline genes being related to each other by insertion/deletion events.
We thus we provide a command line parameter to adjust the Smith-Waterman gap-opening penalty to vary sensitivity to insertion/deletion mutations.
Any candidate insertion/deletion mutations are then flagged, after which the HMM can be run with and without the insertion/deletion.

\paragraph*{Multi-HMMs}
\label{SECmultiHMMs}
As described above, \ham\ is able to do inference under a model which simultaneously emits an arbitrary number of symbols $k$.
When $k=2$ this is typically called a pair HMM \cite{Durbin1998-uq}, and we call the generalized form a multi-HMM ($k \geq 2$).
One can also think of this as doing inference while constraining all of the sequences to come from the same path through the hidden states of the HMM.
In our setting, the $k$ sequences resulting from such a multi-HMM model are the various sequences deriving from a single rearrangement event (which differ only according to point substitution from somatic hypermutation).

We can use this ability to perform improved inference on a collection of sequences deriving from the same rearrangement event, allowing us to integrate out much of the uncertainty due to somatic hypermutation in each single sequence.
In other words, not all mutations are shared between clonally related sequences, and this provides valuable information that is only available if the HMM can emit several sequences simultaneously.

Although this approach allows us to share inferential power among various sequences within the germline regions, with a standard VDJ HMM topology (as in Fig.~\ref{FIGhmmTopology}) the N-region does not contribute to the likelihood in a meaningful way.
In order to allow likelihood contributions from the N-region, we replace the single insert state with four states, corresponding to naive-sequence N-addition of A, C, G, and T (Fig.~\ref{FIGfourInsertStates}).
The emissions of these four states are then treated as for actual germline states (Fig.~\ref{FIGaFewHmmStates}): the A state, for example, has a large probability of emitting an A, and a complementary probability (equal to the observed mutation probability) of emitting one of the other three bases.
For example, if we pass in five sequences that share identical N-regions except that one sequence has a single mutated base, the HMM uses the information from the four non-mutated bases to conclude that the difference in the fifth sequence is likely to be a mutation from the un-mutated ancestral sequence.

\paragraph*{Parameter estimation}
\label{SECmethodsParameterEstimation}
The parameters of our HMM consist of allele usage probabilities, transition probabilities, and emission probabilities; as described above we estimate categorical distributions for each of these.
The inferred distributions of these parameters come from either a previous generation of HMM estimates, or from Smith-Waterman (as described below).

In order to convert from substitution probability $p$ to emission probabilities, we set the probability of emission for each non-germline base to $p/3$, and correspondingly set the probability of germline emission to $1-p$.
In the HMM implementation, it is trivial to account for different probabilities of mutating to different bases (instead of simply using $f/3$ for all three).
However, we do not know of phylogenetic sequence simulation software (we use \textsf{bppseqgen} from Bio++~\cite{bppseqgen}) that implements both this and per-position mutation rates, and thus lacking reliable means of validation for this feature we leave it out.

Transition probabilities are set according to empirical frequencies of inferred exonuclease deletion and N-region lengths.
For example, again sub-setting by allele, the frequency with which we observe a 3' exonuclease deletion of length one gives us the transition probability from the second-to-last state in this allele to the end state (thereby bypassing the last state in the allele).
This logic is repeated for all positions in each allele, and also for the 5' end (where instead of a transition to the end state, it is a transition from the initial state).
Note that we include 5' V and 3' D exonuclease deletions as a convenience (dashed lines in Fig.~\ref{FIGhmmTopology}) to account for varying read lengths.
The N-region state self-transition probability is set to the inverse of the observed mean N-region length in data, and thus N-region lengths are modeled according to a geometric distribution with the correct mean length.
While the choice of a geometric distribution is simply a result of HMMs' inherent lack of memory, in practice we observe that N-region lengths are not far from geometrically distributed (Fig.~\ref{FIGhumanVsHumanInsertionLengths}).

Finally, the allele choice probabilities denoted $q_{g}$ above are simply set to the observed frequency of each allele.

For all parameters, we perform an initial estimation step using the Smith-\hspace{0pt}Waterman based methods of~\cite{ighutil}.
These provisional parameters are used to build a set of HMMs which we then run on the same data in order to obtain a more accurate set of parameters.
We can feed these parameters back into the HMM and continue the process recursively (this is Viterbi training in the language of~\cite{Durbin1998-uq}), but in practice we find no significant improvement after the first iteration.
In future versions we may implement full Baum-Welch parameter estimation~\cite{Durbin1998-uq}.
However, given the level of agreement which we observe for Viterbi training, we do not expect large improvement from moving to full Baum-Welch.

In order to correctly estimate parameter uncertainties we would like to know the sample size: the number of independent rearrangement events.
This is not the same as the number of unique reads, because several reads can stem from a single rearrangement through somatic hypermutation and clonal expansion.
In principle the sample size for each measurement should be the number of observed rearrangement events.
However, we can only observe sequences rather than rearrangement events, and thus do not know precisely how many rearrangement events we have observed.
Previous work has shown that a significant majority of B cell rearrangements are represented by only one sequence in healthy humans, even in a deep sample from the memory compartment~\cite{Bashford-Rogers2013-xv}.
In order to be conservative we posit that B-cells are members of clones with two members, and thus multiply all uncertainty measurements (e.g.\ in Fig.~\ref{FIGhumanVsHumanDeletion}) by a factor of $\sqrt{2}$ corresponding to this assumed two-fold reduction in sample size.

\paragraph*{Parameter merging in small samples via tiered aggregation}
As mentioned above, a more highly parameterized model is both allowed by and requires large data sets.
In order to facilitate the use of \partis\ on data sets of varying size, and also to allow accurate inference of rare clones, we have thus implemented a scheme that automatically reduces the detail of the model on smaller samples.
The basic idea of this scheme, which we call ``tiered aggregation'', is to merge together the information from increasing numbers of germline alleles, in several tiers of similarity, as the sample size decreases.

As shown above, we observe significant parameter variation between alleles, and we thus initially subset all emission and transition probabilities by allele (i.e. use different probabilities for each allele of each gene in V, D, and J).
For large data sets, and for BCR repertoires with relatively homogeneous allele frequency distributions, this approach is sufficient.
If we observe a particular allele only a handful of times, however, we cannot confidently estimate parameters for that allele.
In such cases we thus include information from similar alleles in the parameter estimation.

Specifically, if we observe an allele fewer than $N$ times, we take a simple average over all the observations of its allelic variants when calculating its parameters, rather than just considering that allele.
If we also fail to observe $N$ occurrences of all these allelic variants together, we go on to include all alleles in the same IMGT gene family (e.g. all IGHV1 alleles).
Finally, if with this criterion we still do not reach $N$ observations, we include all alleles from that segment (e.g. all V alleles).
Here $N$ is a tunable parameter, and in practice gives good results in simulation at about 20, which is the value used for the results in this paper.

This procedure thus has the effect of automatically compensating for decreased parameter accuracy in smaller samples by switching to a less parameter-rich model.
We find this strategy to be effective in simulation (Fig.~\ref{FIGhttnBySampleSize}).
As expected, performance decreases with decreasing sample size, although the degradation is minimal for samples larger than a few hundred sequences.
While \partis\ still performs well on fewer than a few hundred sequences, there is little benefit to using it, as one is not taking advantage of its detailed models.
Instead, one could use \ighutil~\cite{ighutil} on small data sets, as in such cases the two will perform comparably in terms of accuracy.

\paragraph*{Validation}
As described above, we find it necessary to validate BCR annotation methods on simulated sequences; we therefore wrote a simulator of rearrangement and somatic hypermutation.
This simulator takes as input a joint categorical distribution over all ``rearrangement values'' which are necessary to specify the rearrangement event (e.g. 3' V exonuclease deletion length, which V allele, etc.).
It then draws a point in this space according to the distribution, and generates a naive sequence according to the coordinates of this point.
N-regions are generated with corresponding length, with N-addition bases chosen randomly according to the empirical nucleotide distribution.
This simulation process implicitly models all correlations between ``rearrangement values,'' even correlations that are ignored in the HMMs.
As an input distribution of rearrangement values to the simulation, we used our inferences on the Adaptive data set.

The resulting unmutated ancestor then diversifies via simulated mutation into a set of sequences in a clonal family.
Specifically, we first use TreeSim~\cite{Stadler2011} to generate a lineage tree descending from this ancestral sequence, with a speciation rate of 1.0 and extinction rate of 0.5, conditioned on the desired number of leaves and on a tree height drawn from the empirical distribution.
The number of leaves is easily configurable in our simulation package either as a constant or as a random variable drawn from some distribution.

The segments are then generated through simulation down this tree using empirical per-position mutation rates, with tree height rescaled by the mean observed ratio of that segment's mutation rate to the rate in all three segments, using the \textsf{bppseqgen} mutation simulator in the Bio++~\cite{bppseqgen} package.
The D segment, for instance, could use a tree height 1.8 times the overall height in a typical human, while the V and J would use 0.98 and 0.87.
N-regions are also mutated, using parameters averaged over the three segments.
This approach yields good agreement between mutation rates in data and simulation (Fig.~\ref{FIGdataVsSimuOverallMutation}).

For simplicity, all plots in this paper use simulation with a constant five leaves per tree; results were consistent when simulating with widely varying constant, as well as non-constant, leaves per tree (results not shown).
The sample for which we show performance comparisons (Table~\ref{TABLEfractionCorrectGenes} and Figs.~\ref{FIGgeneCorrectVsMutFreq},~\ref{FIGhttn}, and~\ref{FIGotherMetrics}) was generated with parameters from human A in the Adaptive data set, except that the mean mutation frequency was doubled (see the original mutation frequencies for the three humans in Fig.~\ref{FIGdataVsSimuOverallMutation}) in order to provide a more challenging inference test.

To check that this simulator generates sequences that are close to real data, we compared its output parameter distributions to the inferred parameters from data (which are the simulator's input).
These distributions are similar (Figs.~\ref{FIGdataVsSimuDeletionInsertion},~\ref{FIGdataVsSimuOverallMutation}, and~\ref{FIGdataVsSimuMutationPerPosition}), showing that the simulator accurately recapitulates the processes generating the true data.
Given this accuracy, the compatibility between true and inferred distributions in Fig.~\ref{FIGotherMetrics} shows that our original parameters inferred from data are indeed close to the true parameters in data.

The validation is run within the \texttt{partis} directory in the main repository using the command \texttt{scons validate}.

We emphasize that this simulator's only inputs are the inferred empirical parameter distributions, and that it is thus entirely independent of \partis' HMM machinery.
\partis's inference framework, meanwhile, is presented with only the simulated sequences; specifically, it has no information about the simulation or its parameters (information which would not be available in a real data set).

\section*{Acknowledgements}
The authors would like to thank
Trevor Bedford,
Lily Blair,
Robert Bradley,
Connor McCoy,
Vladimir Minin,
Steve Quake,
Harlan Robins,
and
Christopher Vollmers
for helpful discussions and for sharing data.
\forarxiv{
This research was funded by National Institutes of Health grant R01 GM113246-01, and in part by a 2013 emerging opportunity award from the University of Washington Center for AIDS Research (CFAR), an NIH funded program under award number P30AI027757 which is supported by the following NIH Institutes and Centers (NIAID, NCI, NIMH, NIDA, NICHD, NHLBI, NIA, NIGMS, NIDDK).
}

\clearpage
\bibliography{hmm-methods-paper}

\clearpage

\beginsupplement

\section*{Supplementary Information}

\FIGhumanVsHumanDeletionVollmers
\FIGhumanVsHumanMutationVollmers
\FIGhumanVsHumanInsertionLengthsVollmers
\FIGmeanVarianceVollmers
\FIGdataVsSimuDeletionInsertion
\FIGdataVsSimuOverallMutation
\FIGdataVsSimuMutationPerPosition

\end{document}